%% file: Paper_August1.tex
\documentclass[preprint,authoryear,12pt]{elsarticle}

 \usepackage{graphicx, subfigure, float}
\usepackage{amssymb}
\usepackage{amsthm}
\usepackage{amsmath}
\usepackage{setspace}
\usepackage{float}
\usepackage{rotating, booktabs}
\restylefloat{table}
\usepackage[utf8x]{inputenc}

\newcommand{\pkg}[1]{{\normalfont\fontseries{b}\selectfont #1}}
\let\proglang=\textsf
\let\code=\texttt

\journal{Computational Statistics \& Data Analysis}

\begin{document}

\begin{frontmatter}


\title{Efficient Optimization of the Likelihood Function in Gaussian Process Modelling}

\author[mun]{A. Butler}
 \ead{adsb85@mun.ca}

  \author[mun]{T.D. Humphries}
 \ead{thumphries@mun.ca}
 \author[acadia]{P. Ranjan}
 \ead{pritam.ranjan@acadiau.ca}
  \author[mun]{R.D. Haynes\corref{cor1}}
 \ead{rhaynes@mun.ca}
\cortext[cor1]{Corresponding author}
\address[mun]{Department of Mathematics and Statistics, Memorial University, St. John's, NL,  Canada,  A1C 5S7 }
\address[acadia]{Department of Mathematics and Statistics, Acadia University, Wolfville, NS,  Canada.  B4P 2R6}

\begin{abstract}
Gaussian Process (GP) models are popular statistical surrogates used for emulating computationally expensive computer simulators. The quality of a GP model fit can be assessed by a goodness of fit measure based on optimized likelihood. Finding the global maximum of the likelihood function for a GP model is typically very challenging as the likelihood surface often has multiple local optima, and an explicit expression for the gradient of the likelihood function is typically unavailable. Previous methods for optimizing the likelihood function (e.g., \cite{P2}) have proven to be robust and accurate, though relatively inefficient. We propose several likelihood optimization techniques, including two modified multi-start local search techniques, based on the method implemented by \cite{P2}, that are equally as reliable, and significantly more efficient. A hybridization of the global search algorithm Dividing Rectangles (DIRECT) with the local optimization algorithm BFGS provides a comparable GP model quality for a fraction of the computational cost, and is the preferred optimization technique when computational resources are limited. We use several test functions and a real application from an oil reservoir simulation to test and compare the performance of the proposed methods with the one implemented by \citet{P2} in the \proglang{R} library \pkg{GPfit}. The proposed method is implemented in a \proglang{Matlab} package, \pkg{GPMfit}.
\end{abstract}


\end{frontmatter}


\input{Intro}
\input{GPmodel}

\input{Method}
\input{Results}

\input{Discuss_Results}

\input{example}

\input{conclusion}

\bibliographystyle{elsarticle-harv}
\bibliography{Bibliography}



\appendix
\section{Test Functions}

\begin{table}[H]
 \centering
 {\footnotesize
\begin{tabular}{|ll|}
 \hline
\textbf{Test Function (d)} & \textbf{Formula y=f(x)}\\
&\\
Hump (1) & $y=1.0316285+4x^2-2.1x^4+\frac{1}{3}x^6$\\
&\\
Goldstein-Price (2) & $y =$\\
& $ [1 + (x_1 + x_2 +1)^2 (19-14x_1+3x_1^2-14x_2+6x_1x_2+3x_2^2)] \times $ \\
& $[30+(2x_1-3x_2)^2(18-32x_1+12x_1^2+48x_2-36x_1x_2+27x_2^2)]$\\
&\\
Schwefel (5) & $y=2094.9 -\sum\limits_{i=1}^5 {x_i \sin{\left(\sqrt{|x_i|}\right)}}$\\
&\\
Hartmann (6) & $y= -\sum\limits_{i=1}^6 {\alpha \cdot \text{exp}[-\sum\limits_{j=1}^6{B_{ij}(x_j-Q_{ij})^2}]}$\\
& $\alpha = [1,1.2,3,3.2]$, \quad  { $B=\begin{bmatrix}
  10 & 3 & 17 & 3.5 & 1.7 & 8 \\
  0.02 & 10 & 17 & 0.1 & 8. & 14 \\
  3 & 3.5 & 1.7 & 10 & 17 & 8 \\
  17 & 8 & 0.05 & 10 & 0.1 & 14
 \end{bmatrix},$}\\
 & $Q=\begin{bmatrix}
 0.1312 &	0.1696 &0.5569& 0.0124 & 0.8283 &0.588\\
0.2329& 0.4135&0.8307 &0.3736 &	0.1004 & 0.9991\\
0.2348 &0.1451& 0.3522& 0.2883 & 0.3047 & 0.6650\\
0.4047 & 0.8828 &0.8732 & 0.5743&0.1091& 0.0381
 \end{bmatrix}$\\
 &\\
Rastrigin (10) & $y=10n + \sum\limits_{i=1}^{10} ({x_i^2-10 \cos{(2 \pi x_i)}})$\\
&\\
Rosenbrock (10) & $y=\sum\limits_{i=1}^{9}[100(x_i^2-x_{i+1})^2+(x_i-1)^2]$\\
&\\
Perm (12) & $y=\sum\limits_{i=1}^{12}\left[\sum\limits_{j=1}^{12}\left[(j^i+0.5)\left(\frac{x_j}{j}\right)^{i-1}\right]^2\right]$\\
\hline

 \end{tabular}
\caption{Test functions and corresponding formula used for evaluating the performance of the $\mathcal{L}_\beta$ optimization process in GP modelling.}
 \label{formula} }
 \end{table}



\end{document}

%% file: Intro.tex
\section{Introduction}
Computer simulators are useful tools for modelling  complex real world systems that are either impractical, expensive, or time consuming to physically observe. For example, the energy generated by the tides of large ocean basins \citep{Tide}, the estimation of the magnetic field generated near the Milky Way \citep{Short}, and the analysis of the flow of oil in a reservoir \citep{Oil} -- the latter of which motivated this research -- can be achieved through the use of computer simulators. That being said, realistic computer simulators can be computationally expensive to run, and as a result are often emulated using statistical models, such as Gaussian Process (GP) models \citep{S}.

The maximum likelihood approach for fitting a GP model to deterministic simulator output requires the minimization of the negative log-likelihood, or deviance ($-2\log(L)$). \citet{Ras} proposed the use of either a randomized multi-start conjugate gradient method or Newton's method for this problem. Explicit information about the gradient of deviance, however, cannot easily be obtained, and the deviance surface often has multiple local optima, making the optimization problem challenging \citep{P2}. Derivative-free optimization techniques, such as the genetic algorithm used by \citet{P1}, or the differential evolution algorithm used by  \citet{Dob}, are robust, but can be computationally inefficient. Gradient approximation methods, such as the Broyden-Fletcher-Goldfarb-Shanno method (BFGS)~\citep{Bro,Flet,Gold,Shan}, are generally faster, but have the potential to converge only locally if poorly initialized. \citet{P2} proposed a clustering-based multi-start BFGS algorithm, which allows for a more global search to be performed. Nonetheless, this method demands several executions of BFGS, which is also computationally expensive.

In this paper we investigate several optimization techniques in order to improve the efficiency of the likelihood optimization process. Each technique is a combination of the global and local search components. We propose using the Dividing Rectangles algorithm (DIRECT)~\citep{Direct} as an alternative to the clustering-based approach for choosing the starting point(s) of BFGS. In terms of the local search, we compare the efficiency of BFGS with Implicit Filtering (IF), a sophisticated pattern search algorithm developed by \citet{IF} for multimodal noisy functions. We use several test functions and a real application from an oil reservoir simulation to compare the performance of different optimization techniques; measured by the prediction accuracy (optimized deviance and root mean squared prediction error) and number of deviance function evaluations (FEs) required to optimize the deviance. After an extensive case study we find that a hybrid approach of DIRECT-BFGS is the most efficient optimization technique for fitting such GP models. 

The remainder of the paper is outlined as follows. Section~$2$ describes the GP model and the main components of the newly developed \proglang{Matlab} package \pkg{GPMfit}. In Section~$3$ we briefly outline the optimization techniques used for the deviance minimization. Section~$4$ provides the results and analysis for several test functions, followed by a real world example in Section~$5$ where the GP model is fit to an oil reservoir simulator using our proposed method. Concluding remarks are provided in Section~$6$. 

%% file: GPmodel.tex
\section{The Gaussian Process Model}

The GP model requires as input a set of design points, $x_i = (x_{i1},...,x_{id})'$, and the corresponding simulator outputs, \ $y_i=y(x_i)$, where $i=1,...,n$, and $n$ is the number of user supplied design points. Here, the prime, $'$, denotes the transpose of vectors and matrices. We assume that the simulator provides a scalar valued output, $y_i$, for each design point $x_i$, and we denote $Y=(y_1,...,y_n)'$ as the $n \times 1$ vector of simulator outputs. The simulator output is modelled as
 \begin{equation*}
 y_i=\mu +z(x_i),
 \end{equation*} where $\mu$ is the overall mean, and $z(x_i)$ is a GP with $E[z(x_i)]=0$, $\text{Var}[z(x_i)]=\sigma^2$, and $\text{Cov}[(z(x_i),z(x_j)]=\sigma^2 R_{ij}$. %

The $n\times n$ spatial correlation matrix $R$ defines the degree of dependency between design points, based on their observed simulator value. Following \citet{P2}, we use the Gaussian correlation matrix, $R$; a special case of the power exponential correlation family defined as
 \begin{equation} \label{Rij}
R_{ij} = \prod \limits_{k=1}^d \text{exp}\{-10^{\beta_k}|x_{ik}-x_{jk}|^{p_k}\} \quad \text{for all} \ i,j.
\end{equation}
Here $p_k =2$ is the smoothness parameter, and $\bf{\beta}$ $= (\beta_1,...,\beta_d)$ is a $1 \times d$ vector of correlation hyper-parameters which measures the sensitivity of the response to the spatial distribution of $|x_{ik}-x_{jk}|^2 $ for all $i, j \in \{1,...,n\}$ and $k \in \{1,...,d\}$ \citep{Lo}.

The formulation of correlation function in Equation (\ref{Rij}) is slightly different than the popular form of Gaussian correlation, which replaces $10^{\beta_k}$ with  $\theta_k$ (e.g., in \citet{P1}). \cite{P2} demonstrate that the deviance surface with $\beta$-parametrization shown in Equation (\ref{Rij}) is much easier to optimize as compared to the commonly used $\theta$-parametrization.

\citet{S} show that the best linear unbiased predictor (BLUP) at a given $x^*$ in the input space (typically normalized to $[0, 1]^d$) is
\begin{equation*}
\begin{aligned}
 \hat{y}(x^*) & =  \hat{\mu}+r'R^{-1}(Y-\boldsymbol{1_n}\hat{\mu}) \\
 & = \left[\frac{(1-r'R^{-1}\boldsymbol{1_n})}{{\boldsymbol{1_n}}'R^{-1}\boldsymbol{1_n}} {\boldsymbol{1_n}'} + r'\right]R^{-1}Y \\
 &\equiv C'Y,
\end{aligned}
\end{equation*}
where $r=[r_1(x^*),...,r_n(x^*)]'$, and $r_i(x^*)=\text{corr}[z(x^*),z(x_i)]$ is the correlation between $z(x^*)$ and $z(x_i)$. The GP model also returns the associated uncertainty estimate, $s^2(x^*)$, as measured by the mean squared error (MSE),
\begin{equation}
s^2(x^*)=E \left[ ( \hat{y}(x^*)- {y}(x^*))^2\right] = \hat{\sigma}^2 (1-2C'r+C'RC).
\label{MSE}
\end{equation}

The model fitting process requires the estimation of $\mu, \sigma^2$ and $\beta$. The closed form estimators of the mean and variance are given by
\begin{equation*}
\hat{\mu}(\beta)=(\boldsymbol{1_n}' R^{-1} \boldsymbol{1_n})^{-1}(\boldsymbol{1_n}' R^{-1}Y)
\end{equation*}
and
\begin{equation*}
 \hat{\sigma}^2(\beta) = \frac{(Y-\boldsymbol{1_n}\hat{\mu}(\beta))'R^{-1}(Y-\boldsymbol{1_n}\hat{\mu}(\beta))}{n},
\end{equation*}
respectively, and are used to obtain the profiled negative log-likelihood or deviance (ignoring the unimportant terms like $\log(\sqrt{2\pi})$ and $\log(n)$):
 \begin{equation} \label{profileL}
 \mathcal{L}_\beta =
 \log{(|R|)} + n\log{[(Y-\boldsymbol{1_n}\hat{\mu}(\beta))'R^{-1}(Y-\boldsymbol{1_n} \hat{\mu}(\beta))]},
 \end{equation}
where $\bf{1_n}$ is an $n \times 1$ vector of all ones. The most difficult part of fitting the GP model is to find
\begin{equation*}
 \beta^* = \text{arg} \ \underset{\beta \in \mathbb{R}^d}{\text{min}} (\mathcal{L}_\beta).
 \end{equation*}

Equation \eqref{profileL} shows that evaluating $\mathcal{L}_\beta$ requires computing both the action of the inverse, $R^{-1}$, and determinant, $|R|$, of the correlation matrix. If any pair of design points are sufficiently close to one another, the matrix $R$ can become near-singular, resulting in unstable computation of $R^{-1}$ and $|R|$. This can lead to an unreliable model fit.

To overcome this instability, we follow a popular technique developed by \citet{S}, \citet{N}, and \citet{Book}, which adds a small ``nugget'' parameter, $\delta$, to the model fitting procedure. The inclusion of $\delta$ smoothes the model predictions, and consequently the GP model fit will no longer exactly interpolate the design points \citep{Oh, Wahba}. To avoid over-smoothing, \citet{P1} introduce a lower bound on the nugget parameter,
\begin{equation*}
\delta_{lb}=\text{max}\left\{\frac{\lambda_n(\kappa(R) -e^a)}{\kappa(R)(e^a-1)},0\right\},
\end{equation*}
where $\kappa(R)=\lambda_n/\lambda_1$ is the $2$-norm condition number and $\lambda_1 \leq \lambda_2 \leq ...\leq \lambda_n$  are the eigenvalues of $R$. That is, $R$ is simply replaced by $R_{\delta} = R + \delta_{lb} I$ in (\ref{profileL}). The over-smoothness can further be reduced by using the iterative regularization technique proposed by \citet{P1}. In our simulations, we scale the input-space domain to $[0,1]^d$, and the design points are generated via a space-filling maximin Latin hypercube design (LHD). The desired threshold for $\kappa(R)$ is $e^a$, where $a=25$ is recommended for a space-filling LHD \citep{Mck}. Under this configuration, $\delta_{lb}$ values remain relatively small, if not zero, and therefore the iterative approach is not needed.

Although there are several choices for the correlation function, we focus on the Gaussian correlation, with $p_k=2$, because of its smoothness property 
and popularity in other areas such as machine learning (radial basis kernels) and geostatistics (kriging). In practice, however, we can increase the stability of inverse and determinant computation by slightly lowering the smoothness parameter, $p_k$, of Equation~\eqref{Rij}, such that $p_k \lessapprox 2$ (e.g., $p_k=1.99$). By setting $p_k=1.99$, the smoothness is not affected significantly and the occurrence of near-singularity is substantially reduced; though not completely resolved as instability may still occur if the design points are extremely close to each other in input space. In this paper, we used $p_k=2$ with $\delta_{lb}$ as given above for all simulated examples in Section~4, and $p_k=1.99$ with the same formula for $\delta_{lb}$ in the oil reservoir application in Section~5.

%% file: Method.tex
\section{Optimization Methodology}

\begin{figure}[htp]
\centering
\includegraphics[height=2.25in]{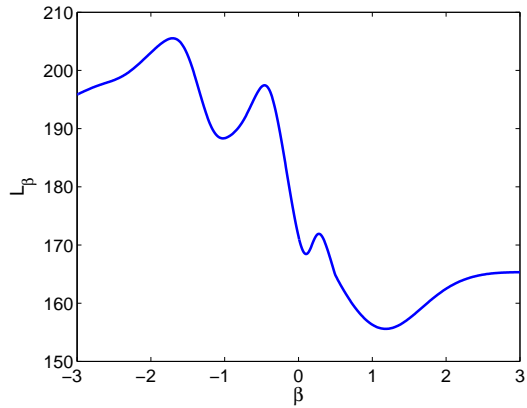}
\caption{ 1-D plot of $ \mathcal{L}_\beta$ surface for the 1-D Hump function and a given set of design points. The Hump function is described in Appendix A.}
\label{like_profile}
\end{figure}

Our objective function, $ \mathcal{L}_\beta$, has a complicated dependency on $\beta$, namely in the form of the mean estimator, $\hat{\mu}(\beta)$,  the correlation matrix, $R$, and the nugget parameter, $\delta$. Thus, it is difficult to extract an explicit gradient, $\nabla \mathcal{L}_\beta$, and the optimization methods that require the user to provide an expression for $\nabla \mathcal{L}_\beta$ are not applicable here. We can, however, compute numerical approximations to $\nabla \mathcal{L}_\beta$, which is implicitly performed in both BFGS and IF algorithms. That said, $\mathcal{L_\beta}$ may contain several local optima and flat regions (see Figure~\ref{like_profile}). Thus, a strictly descent-based optimization approach may not be desirable.

 In this section, we first present a brief description of the local optimization algorithms used herein, namely BFGS and IF. In Sections $3.2$ and $3.3$ we describe the global optimization techniques, namely multi-start clustering and DIRECT. We conclude with a brief discussion on the bound constraints imposed on the $\beta$ search space.

\subsection{BFGS and Implicit Filtering}
The BFGS algorithm is a quasi-Newton optimization technique that uses a rank-two Hessian update formula to locate an optimal $\beta$ \citep{Tool}. At iteration $k$, the algorithm obtains a descent direction, $q^{(k)}$, by approximating the Hessian matrix, $H^{(k)}$, and the gradient, $\nabla \mathcal{L}_\beta^{(k)}$, at a current point, $\beta^{(k)}$, and solving
$$H^{(k)} q^{(k)}= -\nabla \mathcal{L}_\beta^{(k)}\cdot \beta^{(k)}  \quad  \text{for} \quad   k=0,1,... \quad .$$
A line search procedure then determines a suitable step-size, $\alpha^{(k)}$, along $q^{(k)}$, in order to obtain an updated solution, $\beta^{(k+1)}$, given by 
$$ \beta^{(k+1)} = \beta^{(k)} + \alpha^{(k)} q^{(k)}.$$
We use \proglang{Matlab}'s built-in unconstrained optimization routine \code{fminunc} for an implementation of BFGS. We have chosen to use the \textit{medium-scale} implementation of \code{fminunc}, where the user must supply an initial value, $\beta^{(0)}$.

Implicit Filtering (IF) is a sophisticated, deterministic pattern search algorithm designed by \citet{IF} for bound constrained optimization. Specifically, IF is a local optimization algorithm that hybridizes a general pattern search algorithm with a BFGS derivative-approximation algorithm. The pattern search is arranged on a stencil, which, given an incumbent solution $\beta^{(k)}$, will evaluate $\mathcal{L}_\beta$ at $\beta^{(k)} \pm hv_j$, in all coordinate directions $j=\{1,...,d\}$. Here, $v_j = (L_j-U_j)e_j$, where $L_j$ and $U_j$ represent the components of the lower and upper bounds of the search space, respectively, and $e_j$ is the unit coordinate vector. The scale, $h$, varies as the optimization progresses, according to the sequence  $h=\{2^{-m}\}_{m=1}^7$, where $m$ increases by $1$ each time the current stencil fails to find a more optimal position than the incumbent point. As the pattern search progresses, IF constructs a linear least squares interpolant from previously sampled points. After each pattern search phase, the linear interpolant surface is optimized locally using the BFGS algorithm. This process repeats until an optimal $\beta$-parameter is located. The idea is that the pattern search phase, with a suitable step-size, could step over local minima, while the quasi-Newton phase of IF will establish efficient convergence in regions near the global optimum.

The BFGS algorithm, \texttt{fminunc}, does not require user specified bound constraints and works efficiently for smooth objective functions. With appropriate Hessian and gradient approximations and an efficient line search, the BFGS algorithm converges superlinearly to a locally optimal $\beta$-parameter \citep{NL} near the initial solution $\beta^{(0)}$. Conversely, IF does require user specified bound constraints and is designed  for functions that are noisy, with many local optima. The convergence of IF is somewhat slower than superlinear, or equivalently, BFGS \citep{IF}; however, IF is, in principle, more likely to find the global optimum within the bound constraints, due to its pattern search component. The quasi-Newton phase of both algorithms requires computation and storage of an approximate Hessian matrix and solving the resulting system of equations in order to obtain a suitable descent direction, which is generally computationally expensive.

\subsection{Clustering-based multistart technique}
\citet{P2} proposed using a clustering-based multistart BFGS to replace the computationally expensive genetic algorithm used by \citet{P1}. The BFGS algorithm converges more rapidly, but lacks robustness, in that the algorithm has the potential to get stuck in a local minimum, depending on the starting position, $\beta^{(0)}$. \citet{P2} therefore proposed using $2d+1$ starting points of BFGS to improve the chances of global convergence, where $d$ is the dimension of the simulator input (and therefore of $\beta$ as well). These points are determined through sampling and $k$-means clustering method \citep{Mac}, as described below.

\begin{enumerate}
 \item Generate $200d$ $\beta$-vectors within the search space, $S_\beta \subset (-\infty, \infty)^d$ (as defined in Section $3.5$), using a random maximin Latin hypercube design, and evaluate $\mathcal{L_\beta}$ for each $\beta$.
\item From the $200d$ evaluations of $\mathcal{L}_\beta$ obtained from Step~1,  select the $80d$ $\beta$-vectors with the smallest $\mathcal{L_\beta}$ values.
\item Cluster these $80d$ points from Step~2 into $2d$ groups, using the best of $5$ random restarts of $k$-means clustering method. The $2d$ cluster centers will serve as the $2d$ starting points of BFGS.
\item Evaluate $\mathcal{L_\beta}$ at three equidistant points along the main diagonal of the search space, $S_\beta$. The $\beta$-vector with the lowest $\mathcal{L_\beta}$ value is chosen as the $(2d+1)$-th starting point.
\item Begin a run of BFGS from each of the $2d+1$ starting points.
 \end{enumerate}

\begin{figure}[h!]
\centering
\subfigure[$2d+1$ multi-start BFGS.]{\includegraphics[scale=0.45, trim= 5mm 0mm 5mm 5mm, clip=true]{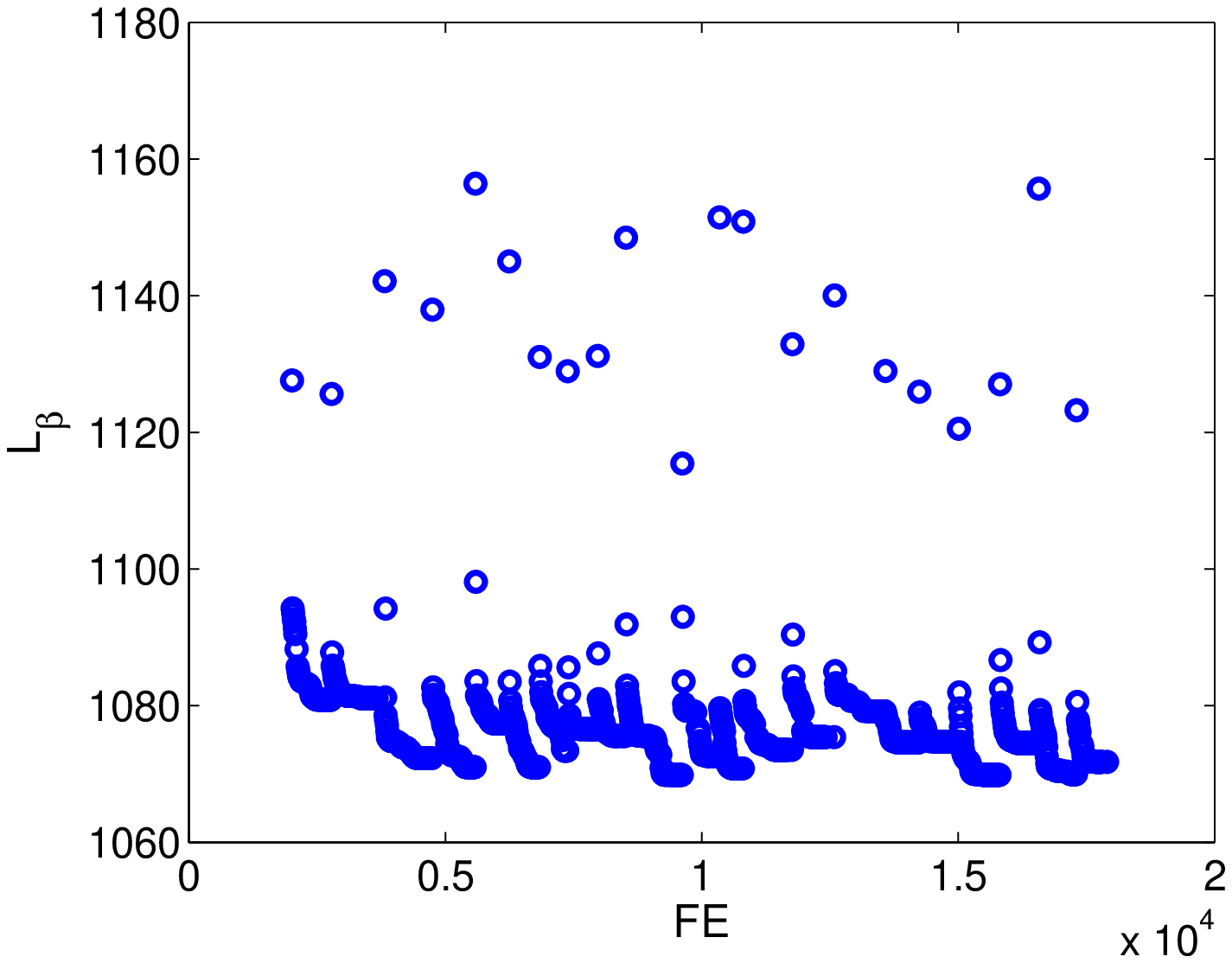}}
\subfigure[$\lceil 0.5d \rceil$ multi-start BFGS.]{\includegraphics[scale=0.45, trim= 5mm 0mm 5mm 5mm, clip=true]{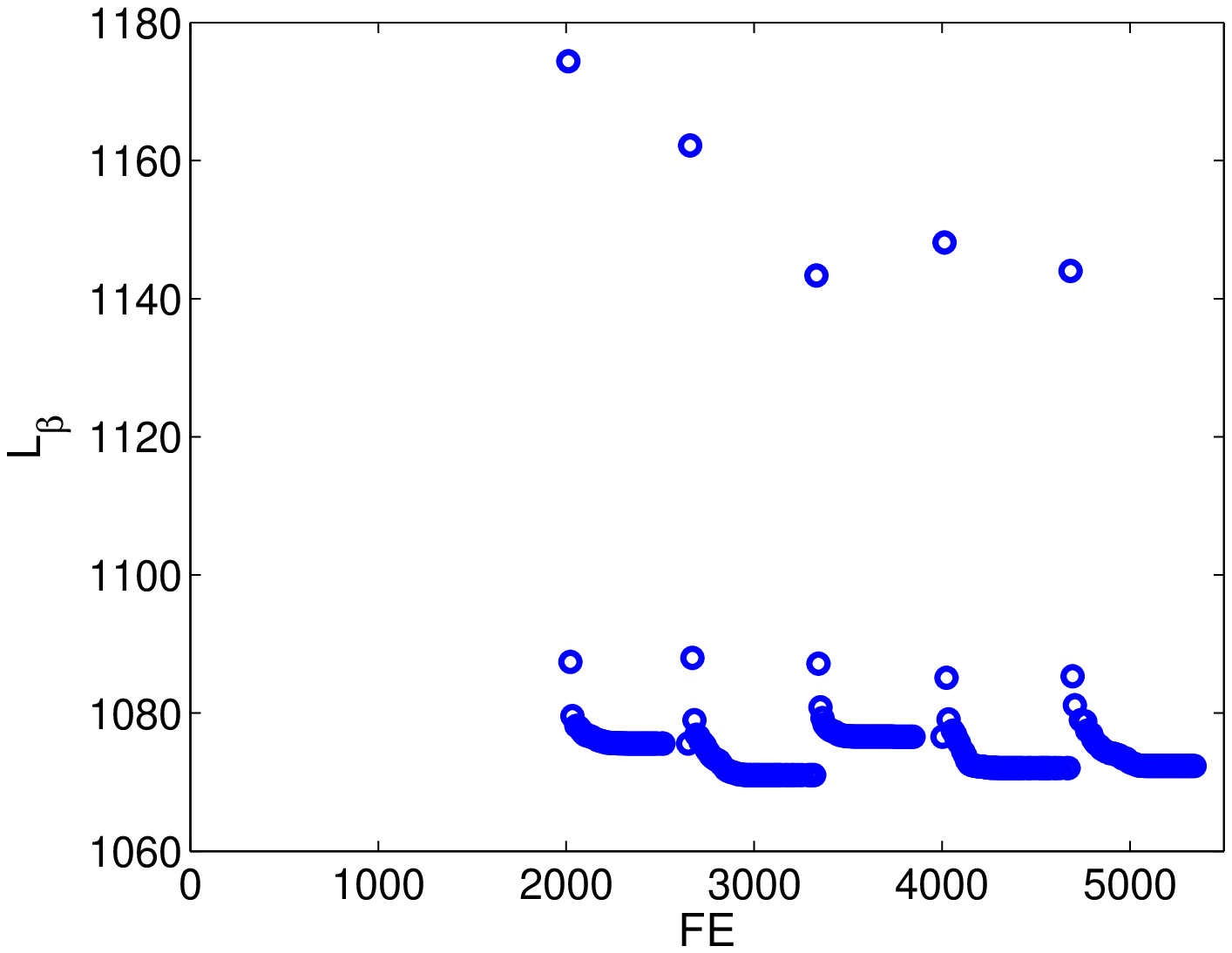}}
\caption[Convergence plot of $2d+1$ BFGS for a hypothetical 10-D function.]{Plot showing the convergence of each BFGS start versus the cumulative number of FEs (includes the initial $200d=2000$ FEs used for clustering), for fitting the 10-D Rastrigin function. The starting points are generated using a random maximin LHD.}
\label{con_plot}
\end{figure}

From thorough experimentation on several test functions, we have observed that executing $2d+1$ starts of BFGS is excessive, and often results in several runs converging to comparable optima. As an example, Figure~\ref{con_plot} shows the convergence of  $2d+1$ multistarts of BFGS (left plot) and $\lceil 0.5d \rceil$ multistarts of BFGS (right plot) as a function of the number of deviance function evaluations (FE), for the problem of fitting a GP model to the 10-D Rastrigin test function (described in Appendix A). It is apparent that three of the five BFGS runs shown in the right plot converge to a value that is comparable to the best value found out of twenty-one BFGS runs in the left plot. We therefore propose reducing the number of cluster centers to $\lceil 0.5d \rceil$ and eliminating the additional starting point obtained from sampling along the main diagonal. We implement two clustering-based techniques: (i) $\lceil 0.5d \rceil$ multistarts of BFGS (the same method as in \cite{P2}), and (ii) $\lceil 0.5d \rceil$ multistarts of IF.

The simulation results (in Table \ref{perform} and Figures \ref{con_plot} and \ref{performance}) show that the $\mathcal{L}_\beta$ value returned by BFGS and IF do not usually change significantly after the first few iterations. Therefore, performing a full search (until the stopping criterion is satisfied) from each starting point $\beta^{(0)}$ may be excessive as well. We therefore investigate a two stage multistart IF method, denoted by IF-2. The process starts with a clustering-based, $\lceil 0.5d \rceil$ multistart IF approach, where each run of IF is limited by a budget of $20d$ FEs. In the second stage, the single run of IF that returns the lowest $\mathcal{L}_\beta$ value will then run to completion.

\subsection{DIRECT Hybrid Techniques}
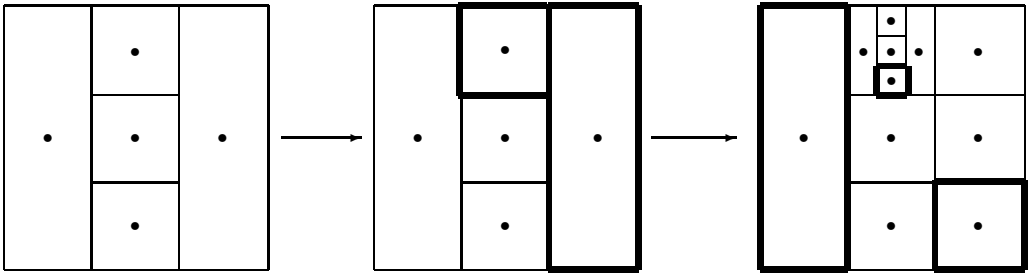
\begin{figure}[h!]
 \begin{center}
\begin{picture}(390,100)(0,0)
  \put(0,0){\line(0,1){100}}
 \put(0,0){\line(1,0){100}}
 \put(100,0){\line(0,1){100}}
 \put(0,100){\line(1,0){100}}

\put(33,0){\line(0,1){100}}
\put(66,0){\line(0,1){100}}
\put(33,33){\line(1,0){33}}
\put(33,66){\line(1,0){33}}

\put(16.5,50){\circle*{3}}
\put(49.5,16.5){\circle*{3}}
\put(49.5,82.5){\circle*{3}}
\put(49.5,50){\circle*{3}}
\put(82.5,50){\circle*{3}}
\put(105,50){\vector(1,0){30}}

  \put(140,0){\line(0,1){100}}
 \put(140,0){\line(1,0){100}}
 \put(240,0){\line(0,1){100}}
 \put(140,100){\line(1,0){100}}

\put(173,0){\line(0,1){100}}
\put(206,0){\line(0,1){100}}
\put(173,33){\line(1,0){33}}
{\linethickness{0.7mm}\put(172,66){\line(1,0){33}}}
{\linethickness{0.7mm}\put(172,66){\line(0,1){33.5}}}
{\linethickness{0.7mm}\put(172,100){\line(1,0){33}}}
\put(156.5,50){\circle*{3}}
\put(189.5,16.5){\circle*{3}}
\put(189.5,83){\circle*{3}}
\put(189.5,50){\circle*{3}}
\put(224.5,50){\circle*{3}}
{\linethickness{0.7mm}
\put(206,0){\line(1,0){34}}
\put(206,100){\line(1,0){34}}
\put(206,0){\line(0,1){100}}
\put(240,0){\line(0,1){100}}
}

\put(245,50){\vector(1,0){32}}

  \put(286,0){\line(0,1){100}}
 \put(286,0){\line(1,0){100}}
 \put(386,0){\line(0,1){100}}
 \put(286,100){\line(1,0){100}}

\put(319,0){\line(0,1){100}}
\put(352,0){\line(0,1){100}}
\put(319,33){\line(1,0){33}}
\put(319,66){\line(1,0){33}}
\put(352,33){\line(1,0){34}}
\put(352,66){\line(1,0){34}}

\put(302.5,50){\circle*{3}}
\put(335.5,16.5){\circle*{3}}
\put(335.5,82.5){\circle*{3}}
\put(335.5,50){\circle*{3}}
\put(368.5,16.5){\circle*{3}}
\put(368.5,50){\circle*{3}}
\put(368.5,82.5){\circle*{3}}

{\linethickness{0.7mm}\put(352,0){\line(1,0){34}}}
{\linethickness{0.7mm}\put(352,0){\line(0,1){33.5}}}
{\linethickness{0.7mm}\put(352,33.5){\line(1,0){33.5}}}
{\linethickness{0.7mm}\put(386,0){\line(0,1){33.5}}}

{\linethickness{0.7mm}\put(286,0){\line(1,0){33}}}
{\linethickness{0.7mm}\put(286,0){\line(0,1){100}}}
{\linethickness{0.7mm}\put(319,0){\line(0,1){100}}}
{\linethickness{0.7mm}\put(319,0){\line(0,1){100}}}
{\linethickness{0.7mm}\put(286,100){\line(1,0){33}}}

\put(330,66){\line(0,1){34}}
\put(341,66){\line(0,1){34}}
\put(330,76.5){\line(1,0){11}}
\put(330,88.5){\line(1,0){11}}

\put(325,82.5){\circle*{3}}
\put(346,82.5){\circle*{3}}
\put(335.6,71.5){\circle*{3}}
\put(335.5,94){\circle*{3}}
{\linethickness{0.7mm}\put(330,66){\line(1,0){11.1}}}
{\linethickness{0.7mm}\put(342,66){\line(0,1){11}}}
{\linethickness{0.7mm}\put(330,77){\line(1,0){11}}}
{\linethickness{0.7mm}\put(330,66){\line(0,1){11}}}

 \end{picture}
\end{center}
\caption{A few iterations of the Dividing Rectangles (DIRECT) algorithm. Bolded rectangles are identified as being potentially optimal and are divided in the following iteration. The \textbullet's  denote the sampled points.}
\label{DIRECT}
\end{figure}

Dividing Rectangles (DIRECT) is a derivative-free, block partitioning algorithm that sequentially samples points in the search space and partitions the domain into hyper-rectangles based on the objective function value (here, $\mathcal{L}_\beta$) at the sampled points. Hyper-rectangles are then identified as being potentially optimal if they contain a sampled point whose function value is more optimal than the sampled points contained by all other hyper-rectangles of equal size. Each potentially optimal hyper-rectangle is then divided into thirds along its longest dimension, and the process repeats. Figure \ref{DIRECT} provides a 2-dimensional visualization of how DIRECT samples and divides its search space. The left panel in Figure~\ref{DIRECT} shows the initial sampling phase and partitioning of the domain. The bold rectangles in the middle panel of Figure~\ref{DIRECT} are identified as being potentially optimal in the following iteration. The rectangles are then partitioned in turn (rightmost figure), and three new rectangles are identified as potentially optimal. The alternating process of partitioning and then identifying potentially optimal rectangles continues, until a stopping criterion is met. See \citet{Direct} for more details.

DIRECT is specifically designed as a global optimization approach, however, since it provides such a thorough exploration of search space it can
be slow to converge locally. We therefore use DIRECT to provide a single, somewhat optimized starting point, $\beta^{(0)}$, from which to begin a 
run of either BFGS or IF, thereby eliminating the need for a multi-start approach. For ease of comparison with clustering-based approach, we provide a budget of $200d$ FEs to DIRECT. 
\newline

 \subsection{Boundaries of the Search Space}

 Many optimization techniques, including IF and DIRECT, require user-supplied bound constraints on the optimization parameters, $\beta$. Although the exact position of the global optimum in $\beta$-space is unknown, we can determine a region, $S_\beta$, where the optimum is likely to be found. \citet{P2} use the structural form of the correlation matrix $R$ to provide an approximate bound for $R_{ij}$. Specifically
 \begin{equation*}
 \text{exp}(-5) \approx 0.0067 \leq R_{ij} \leq 0.9999 \approx \text{exp}(-10^{-4}),
 \end{equation*}
 or equivalently,
 \begin{equation*}
 10^{-4} \leq \sum \limits_{k=1}^d 10^{\beta_k} |x_{ik}-x_{jk}|^2 \leq 5.
 \end{equation*}

Since the assumed input-space is $[0,1]^d$, and the design points are generated via a maximin LHD, the approximate spatial distribution of the $10d$ design points that we use is at most $|x_{ik}-x_{jk}| \approx 1/10$ in any dimension $k$. Moreover, if we assume that the simulator function is equally smooth in all coordinate directions, i.e., $\beta_1\approx \beta_2 \approx \cdots\beta_k$, then the set of $\beta$ values that is likely to contain the global optimum is
 \begin{equation*}
 S_\beta = \{(\beta_1,...,\beta_d): -2-\log_{10}(d) \leq \beta_k \leq \log_{10}(500) - \log_{10}(d), k=1,...,d \}.
 \end{equation*}
The starting points, $\beta^{(0)}$, determined using either clustering or DIRECT, will be confined to the region $S_\beta$.

BFGS is an unbounded optimization algorithm and does not require any user-specified bound constraints on the optimization variables, whereas IF does. The step-size calculated in the pattern search phase of IF is proportional to the physical size of the user supplied bound constraints, which we denote by $S_\beta^{IF}$. From our experimentation on the test functions, we have noticed that if the size of $S_\beta^{IF}$ is too small, the optimization efficiency is compromised as IF requires a large number of FEs in order to converge to an optimal $\beta$-parametrization. Conversely, if $S_\beta^{IF}$ is too large, IF has the tendency to ``jump'' around the potentially optimal regions of $\mathcal{L}_\beta$, resulting in convergence to a suboptimal $\beta$ value.

\citet{P2} note that the optimal $\beta$ values are rarely large and positive, and hence, we impose bound constraints on IF in which the negative $\beta$ region occupies a larger portion of the domain, i.e.,
 \begin{equation*}
 S_\beta^{IF} = \{(\beta_1,...,\beta_d): d(-2-\log_{10}(d)) \leq \beta_k \leq \log_{10}(500), k=1,...,d \}.
 \label{bound}
 \end{equation*}

We acknowledge that the $\beta$-value that globally minimizes $\mathcal{L_\beta}$ may occasionally be positioned outside the provided bounds, $S_\beta$ and $S_\beta^{IF}$. Therefore, included in the \pkg{GPMfit} package is the option to multiplicatively expand or contract the default bound constraints.

%% file: Results.tex
\section{Simulation Results}

We use seven test functions, with input dimensions varying from $d=1$ to $d=12$ to compare the performance of the different optimization techniques discussed in Section~3. The formulae for all the test functions are provided in Appendix~A. The performance of each optimization technique is averaged over $25$ simulations. For each simulation, $10d$ training design points ($x_i$) and $100d$ validation prediction points ($x_i^*$) are chosen in $[0, 1]^d$. The initial sample points in $\beta$-space for the clustering procedure are randomly generated in $S_{\beta}$ using the Latin hypercube design. All simulations were performed using 64-bit \proglang{Matlab} 2012(b) on a Gentoo Linux operating system with a Core~2 Quad Xeon processor.

 \subsection{Optimization Accuracy}
Recall that our objective is to minimize $\mathcal{L}_\beta$. Typically, the parameter estimate that corresponds to the smallest $\mathcal{L}_\beta$ will provide the most accurate model fit, as measured by the average relative root mean square prediction error (RMSPE) between the GP model fit and the true simulator (test function) response. That is,
\begin{equation}
RMSPE=\sqrt{\sum\limits_{i=1}^{N}(y_i - \hat{y}_i)^2 \biggl/ \sum\limits_{i=1}^N y_i^2}, \quad \textrm{where}\ N = 100d.
\end{equation}

We note that in most real applications, the true RMSPE values cannot be calculated, as the true simulator outputs are unknown at the validation points. One can, however, use the average MSE estimates (see Equation \eqref{MSE}) for performance comparison. The consistency of RMSPE values over $25$ simulations is measured by the standard error in the RMSPE value, given by
\begin{equation}
\text{Std. Err.} = \sigma_{RMSPE}/\sqrt{25},
\end{equation}
where $\sigma_{RMSPE}$ denotes the standard deviation of the RMSPE values. A standard error of one order of magnitude less than the corresponding RMSPE value indicates that our results are fairly consistent over all $25$ simulations.

\subsection{Convergence Efficiency}
We measure the efficiency of an optimization technique by the number of likelihood function evaluations (FEs) required for optimization of $\mathcal{L}_\beta$. Using \proglang{Matlab}'s profiler, we determined that evaluating $\mathcal{L}_\beta$ constituted the bulk of the computational load for all optimization techniques considered. In particular, we determined that computation of the correlation matrix, $R_{\delta}$, demands anywhere from $60\%-90\%$ of the total computation time, depending on the simulator input dimension, $d$. Computation of $R_{\delta}$ is nested within the calculation of $\mathcal{L}_\beta$ and is evaluated once per FE. Therefore, the number of FEs, which is not affected by issues like server load, can be used in place of computation time as a fair measure of optimization efficiency.

\subsection{Discussion}
Table \ref{perform} summarizes the accuracy ($\mathcal{L}_\beta$ and RMSPE) and efficiency (FE) of each optimization technique for six of the seven test functions, namely the {2-D} Goldstein-Price function, the 5-D Schwefel Function, the 6-D Hartmann function, the 10-D Rastrigin function, the 10-D Rosenbrock function and the 12-D Perm Function (see Appendix A for closed form expressions). Results for fitting the 1-D Hump function are not shown, as the performance of all the techniques was essentially the same for this simple test function. The \%$\Delta$ notation in Table \ref{perform} denotes the percent relative difference between value of the performance measure returned by a given technique and the best value found among all techniques. The standard errors are not included in this table; we found that for all cases the standard error was indeed at least one order of magnitude less than the corresponding RMSPE, suggesting that each technique is consistently able to provide the same GP model quality.

\begin{table}[h!]\centering
{\footnotesize
\begin{tabular}{lccrccr}
\toprule
    \textbf{Algorithm} & \multicolumn{3}{c}{\textbf{Goldstein-Price (2-D)}} & \multicolumn{3}{c}{\textbf{Schwefel (5-D)}}\\
    &\%$\Delta \mathcal{L}_\beta$& \%$\Delta$RMSPE & FE
     & \%$\Delta \mathcal{L}_\beta$& \%$\Delta$RMSPE & FE  \\
     \cmidrule(r){2-4} \cmidrule{5-7}
$\lceil 0.5d \rceil$ BFGS  &0.017& $0.201$& \underline{439}&0.100& $2.885$ & 1859\\
$2d+1$ BFGS   &$-$& $3.213$ & 653&$-$& $2.885$& 4277 \\
$\lceil 0.5d \rceil$ IF  &0.007& $-$& 518 &0.206&$0.962$  & 2381\\
 $2d+1$ IF &$-$& $3.213$ & 995  &0.074& $2.404$ & 5735\\
$\lceil 0.5d \rceil$ IF-2 & 0.007& $-$& 546 &0.272& $1.442$ & 1677\\
DIRECT-BFGS& $-$& $3.213$& 449  &0.232& $2.885$ & \underline{1296}\\
%
DIRECT-IF  &$-$& $3.213$ & 498 & 0.243&$-$ & 1304\\
\midrule
     & \multicolumn{3}{c}{\textbf{Hartmann (6-D)}} & \multicolumn{3}{c}{\textbf{ Rastrigin (10-D)}}\\
    & \%$\Delta \mathcal{L}_\beta$&  \%$\Delta$RMSPE & FE
     & \%$\Delta \mathcal{L}_\beta$&  \%$\Delta$RMSPE & FE  \\
    \cmidrule(r){2-4} \cmidrule{5-7}
$\lceil 0.5d \rceil$ BFGS   &0.034& $-$ & 2068 &0.071& $3.865$  & 5553\\
$2d+1$ BFGS  &0.102& $1.064$ & 5526&$-$& $3.865$ & 17853\\
$\lceil 0.5d \rceil$ IF  &1.703& $6.991$ & 2293 &0.624& $-$& 4346\\
 $2d+1$ IF&$-$& $0.304$& 7497 &0.134& $2.899$ & 17284\\
$\lceil 0.5d \rceil$ IF-2 &1.223& $8.511$  & 1866&0.545& $0.483$  & 4011\\
DIRECT-BFGS&0.207& $1.216$ & \underline{1526}&0.255& $1.932$ & \underline{2682}\\
%
DIRECT-IF  &0.207& $1.216$ & 1533& 0.287& $1.932$ & 2774\\
\midrule
     & \multicolumn{3}{c}{\textbf{Rosenbrock (10-D)}} & \multicolumn{3}{c}{\textbf{Perm (12-D)}}\\
    & \%$\Delta \mathcal{L}_\beta$&  \%$\Delta$RMSPE & FE
     & \%$\Delta \mathcal{L}_\beta$&  \%$\Delta$RMSPE & FE  \\
   \cmidrule(r){2-4} \cmidrule{5-7}
$\lceil 0.5d \rceil$ BFGS  &$-$& $-$  & 4562&0.003& 1.370 & 8170\\
$2d+1$ BFGS  &$-$& $-$ & 16245&$-$& 1.643  & 27622\\
$\lceil 0.5d \rceil$ IF  &$-$& $-$ & 6654&0.010& $-$ & 12411\\
$2d+1$ IF & $-$& $-$  & 24725&0.001& 2.055 & 44375\\
$\lceil 0.5d \rceil$ IF-2&0.031& 3.196 & 3775&0.025& $-$& 4935\\
DIRECT-BFGS& 0.003& 0.457 & \underline{2332}&0.011& 0.545 & \underline{3338}\\
DIRECT-IF   & 0.003& 0.457 & 2654&0.012& 0.685 & 3459\\
    \bottomrule

\end{tabular}
}
\caption{Performance comparison of different likelihood optimization techniques on six test functions. Dashed values in the \%$\Delta \mathcal{L}_\beta$ and  \%$\Delta$RMSPE columns indicate that the best overall value was found by this algorithm. Underlined values in the FE column indicate the smallest number of FEs required by any algorithm.}
\label{perform}
\end{table}

%% file: Discuss_Results.tex
The results in Table \ref{perform} show that the $\lceil 0.5d \rceil$ multi-start techniques and DIRECT-based
 techniques provide efficient and reliable alternatives to the $2d+1$ BFGS technique. We observe that the $\lceil 0.5d \rceil$ multi-start and DIRECT-based techniques require anywhere from $20\%$ to $90\%$ fewer FEs than the $2d+1$ BFGS technique (depending on the dimension of the test function), while maintaining a comparable level of optimization accuracy. The relative difference in the $\mathcal{L}_\beta$ value returned by each technique is typically  less  than $1 \%$. As a result, each optimization technique provides comparable GP model quality, as measured by the RMSPE value. Table \ref{perform} shows, however, that for all test functions, the $\lceil 0.5d \rceil$ multi-start IF-2 technique returns a larger average $\mathcal{L}_\beta$ value and requires anywhere from $10\%$ to $50\%$ more FEs than both of the DIRECT-based methods. As a result, IF-2 is not included in the \pkg{GPMfit} package, as more accurate and efficient techniques are clearly available.

The results in Table~\ref{perform} suggest that a slightly suboptimal $\mathcal{L}_\beta$ value can result in an equal, or even slightly better GP model qualityas measured by the RMSPE. For example, for fitting the 10-D Rastrigin function, DIRECT-BFGS provides an RMSPE value that is $1.93\%$ smaller than the RMSPE value returned by $2d+1$ BFGS, despite converging to a $\mathcal{L}_\beta$ value that is sub-optimal by $0.255\%$. This non-monotonic relationship between optimal $\mathcal{L}_\beta$ and GP model quality is presented in a recent paper by \citet{KL}, who argue that one can maintain the quality of the GP model even with a slightly suboptimal $\mathcal{L}_\beta$ value. Furthermore, \citet{Over} suggest that,  due to the difficulties in finding optimal $\beta$-parameters, particularly when the training data $(x_i,y_i)$ is sparse, GP models can be prone to overfitting, which can lead to larger than expected RMSPE values. 
Motivated by this non-monotonic relationship, our goal is to efficiently determine a sufficiently optimal $\mathcal{L}_\beta$ value, without compromising the resulting GP model quality.

\begin{figure}[ht!]
\begin{center}
\subfigure[2-D Goldstein-Price.]{\includegraphics[height=47mm, trim= 1mm 0mm 5mm 5mm, clip=true]{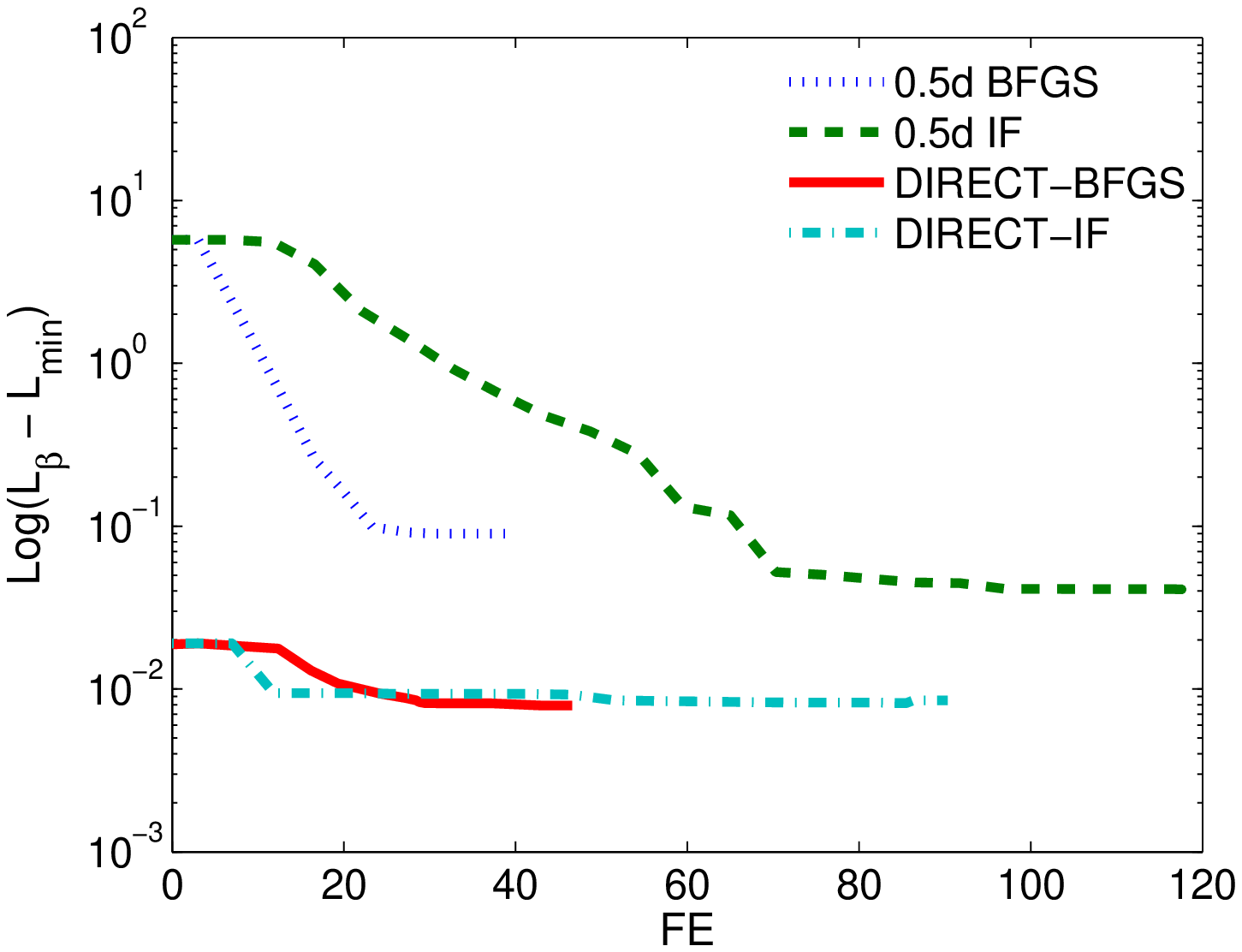}}
\subfigure[5-D Schwefel.]{\includegraphics[height=47mm, trim= 1mm 0mm 5mm 5mm, clip=true]{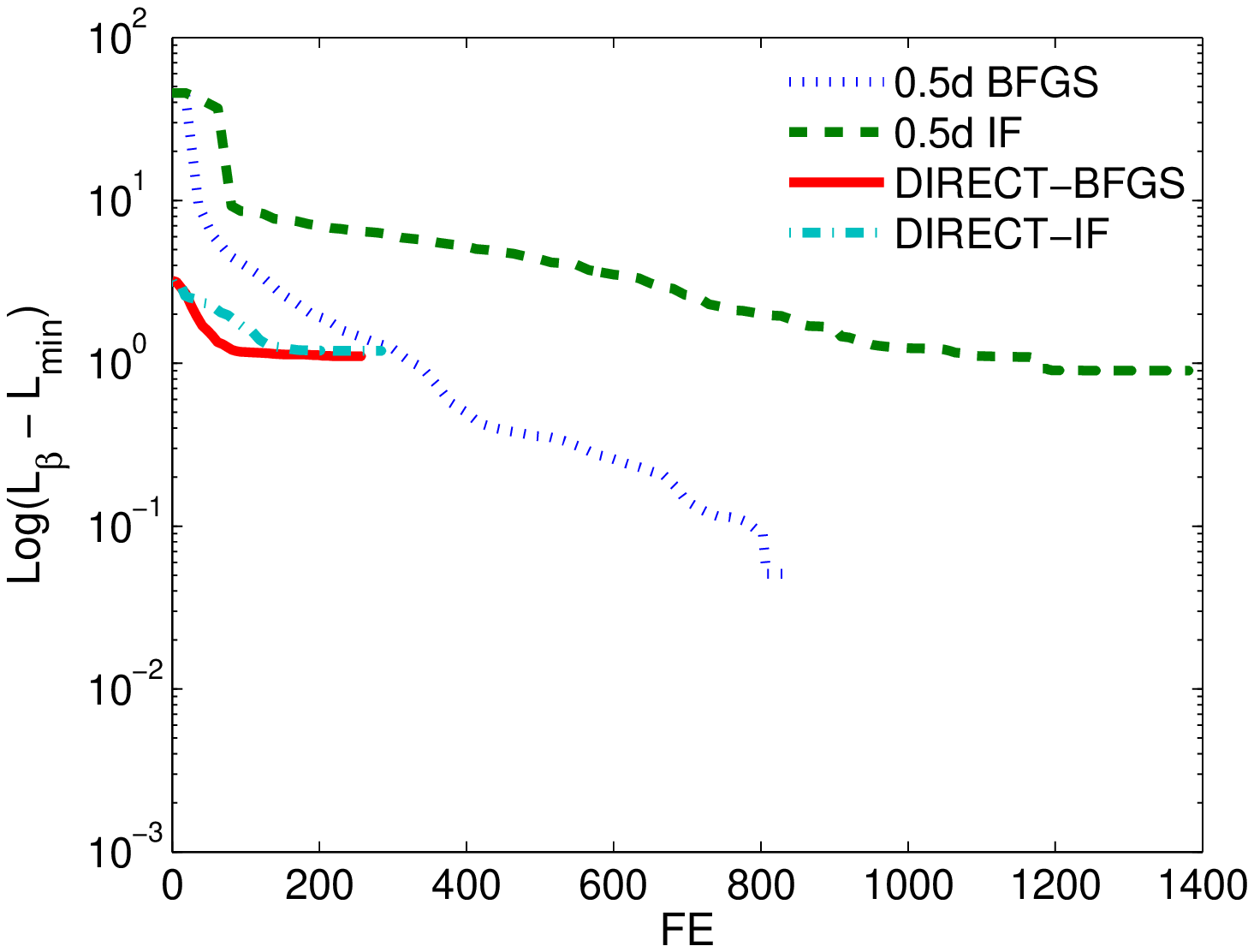}}
\subfigure[6-D Hartmann.]{\includegraphics[height=47mm, trim= 1mm 0mm 5mm 5mm, clip=true]{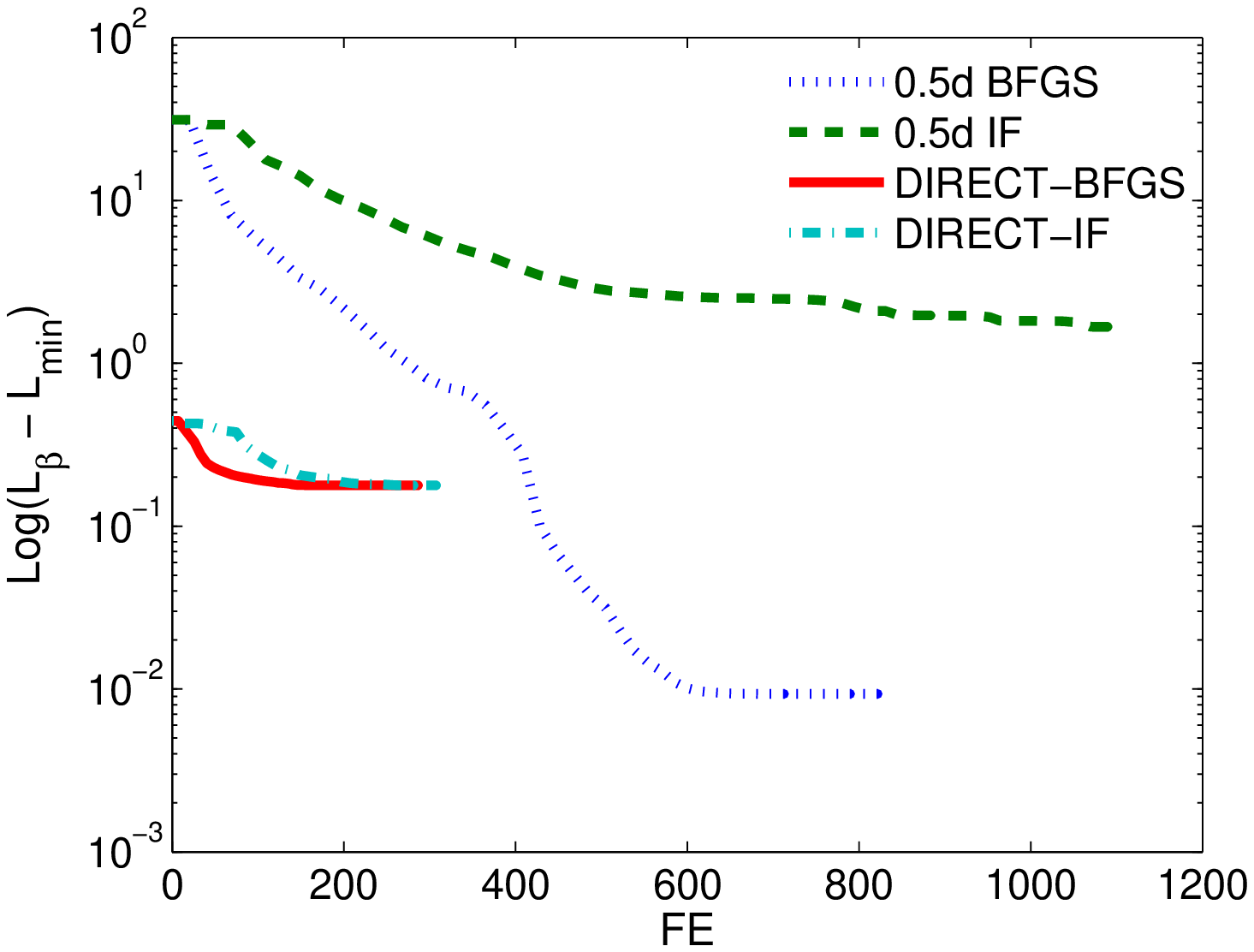}}
\subfigure[10-D Rastrigin.]{\includegraphics[height=47mm, trim= 1mm 0mm 5mm 5mm, clip=true]{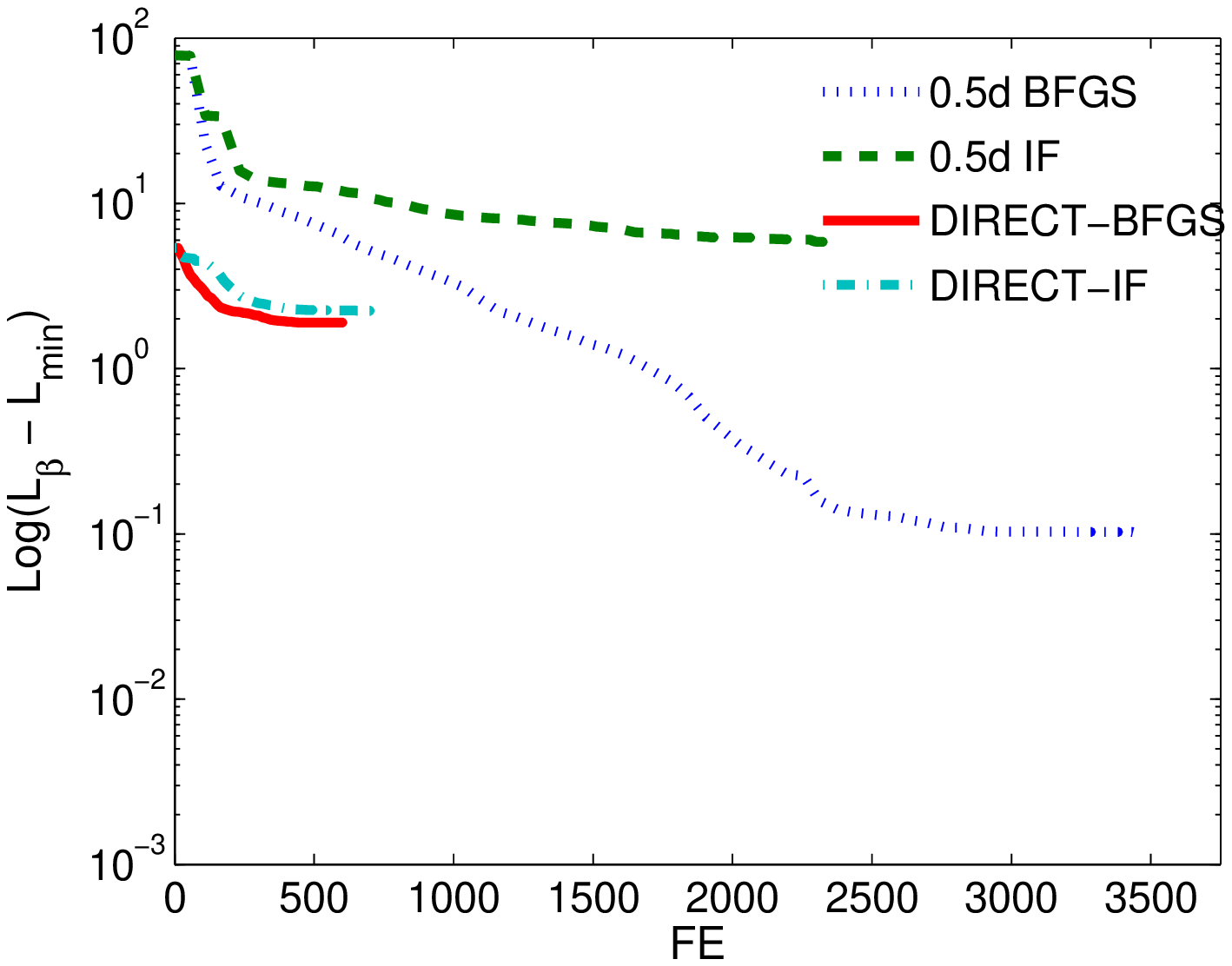}}
\subfigure[10-D Rosenbrock.]{\includegraphics[height=47mm, trim= 1mm 0mm 5mm 5mm, clip=true]{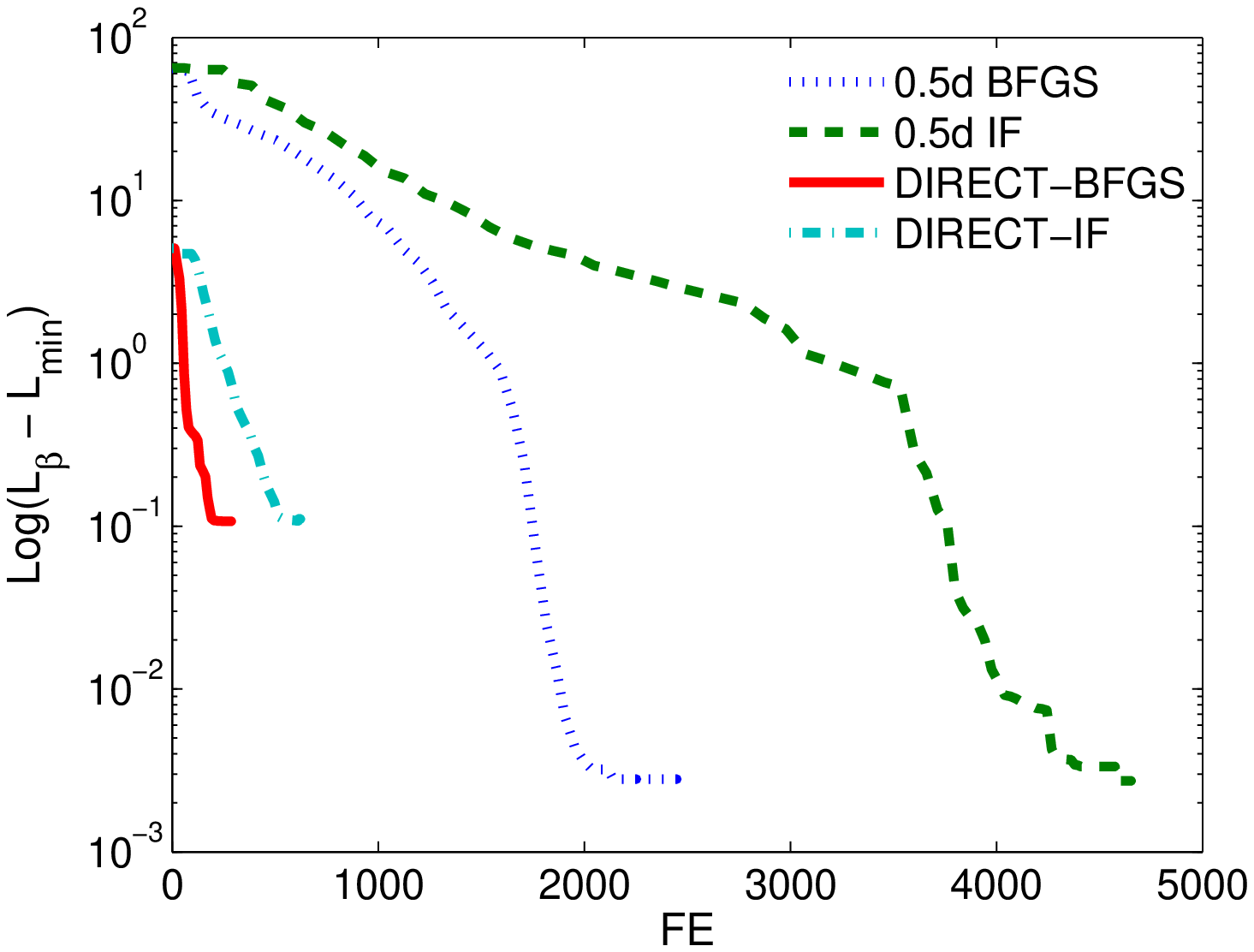}}
\subfigure[12-D Perm.]{\includegraphics[height=47mm, trim= 1mm 0mm 5mm 5mm, clip=true]{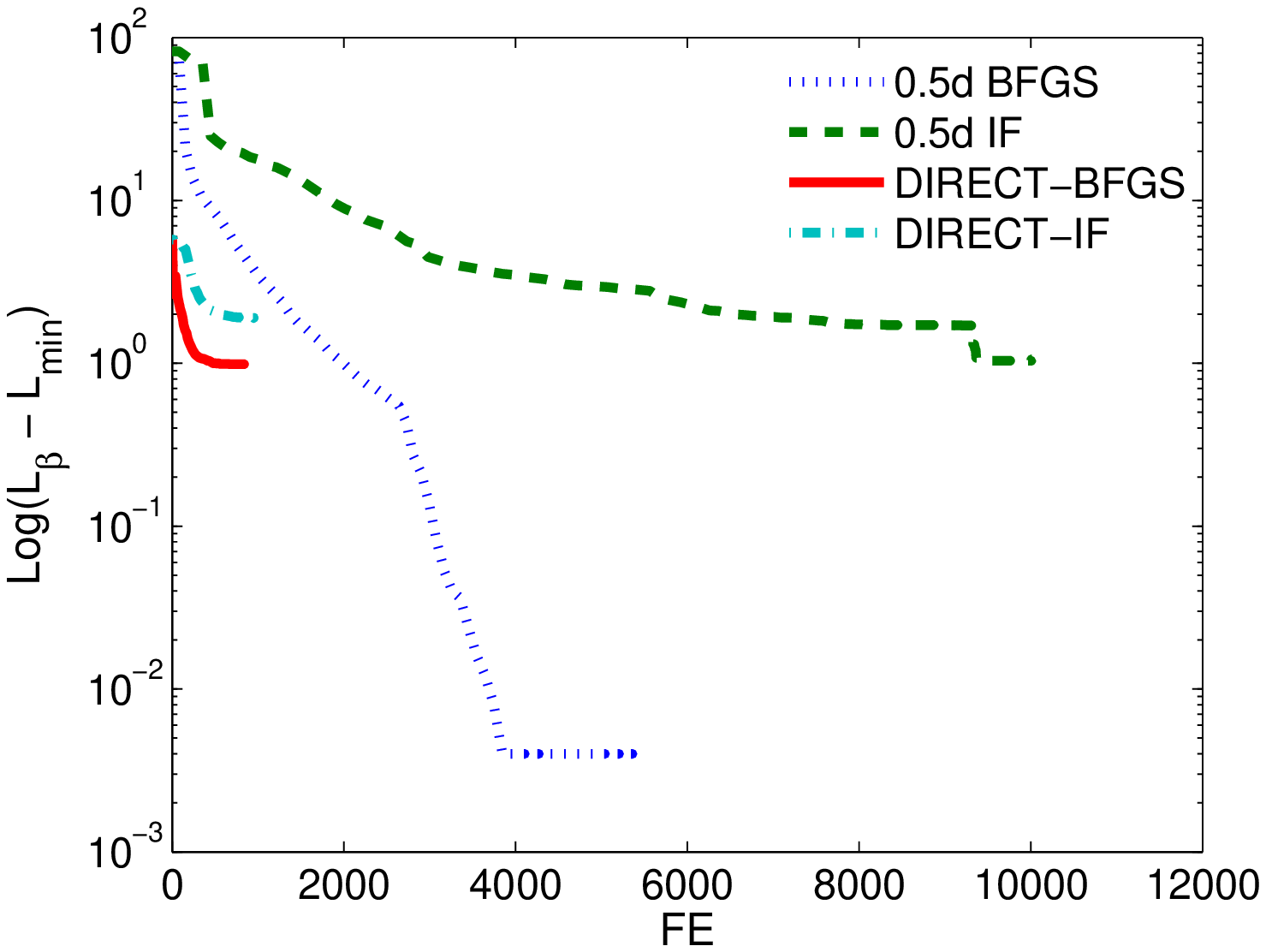}}
\caption[Comparing optimization performance of multi-start and DIRECT methods.]{Semi-log plots comparing the $\mathcal{L}_\beta$ optimization performance of the single start DIRECT-BFGS and DIRECT-IF techniques and the multi-start $\lceil0.5d \rceil$ BFGS and IF techniques, averaged over $25$ simulations.}\label{performance}
\end{center}
\end{figure}

Figure \ref{performance} shows the convergence performance of both the BFGS and IF algorithms after the initial $\beta^{(0)}$ point(s) are determined using either clustering or DIRECT. For the multi-start clustering-based techniques, the convergence plots are displayed as though each run of BFGS or IF were implemented in parallel, from which the best of the $\lceil 0.5d \rceil$ $\mathcal{L}_\beta$ values is plotted versus the cumulative number of FEs. The optimization performance for each technique has been averaged over all $25$ simulations, and is plotted on a semi-log scale as the absolute difference between $\mathcal{L}_\beta$ and the minimum $\mathcal{L}_\beta$ value, $\mathcal{L}_{\text{min}}$, (rounded down to the nearest $10^{-2}$) determined by one of the four techniques.

We first observe that, for both the DIRECT and multi-start approaches, BFGS is more efficient than IF, and converges to a more optimal value of $\mathcal{L}_\beta$. This suggests that BFGS is generally better suited to this optimization problem than IF. Secondly, with the exception of the 2-D test case (Figure \ref{performance}(a)), the $\lceil 0.5d \rceil$ BFGS technique converges to the best $\mathcal{L}_\beta$ value. In general, however, the difference between the $\mathcal{L}_\beta$ value returned by the $\lceil 0.5d \rceil$ BFGS technique and DIRECT-based methods is on the order of $10^{-1}$ to $10^0$, which has only a small effect on the resulting quality of the GP model, as measured by the RMSPE. Moreover, Figure \ref{performance} shows that once the starting $\beta^{(0)}$ point(s) have been determined, the DIRECT-based methods require approximately $\frac{1}{ \lceil 0.5d \rceil}$ as many FEs as the multi-start clustering techniques; this represents a substantial increase in optimization efficiency, particularly as $d$ increases. Thus, the DIRECT-based approaches are able to find an optimum that is only slightly worse than those found by the significantly more expensive clustering-based approaches.

\begin{figure}[htbp]
\begin{center}
\subfigure[Semi-log Convergence Plot.]{\includegraphics[scale=0.40, trim= 1mm 0mm 5mm 5mm, clip=true]{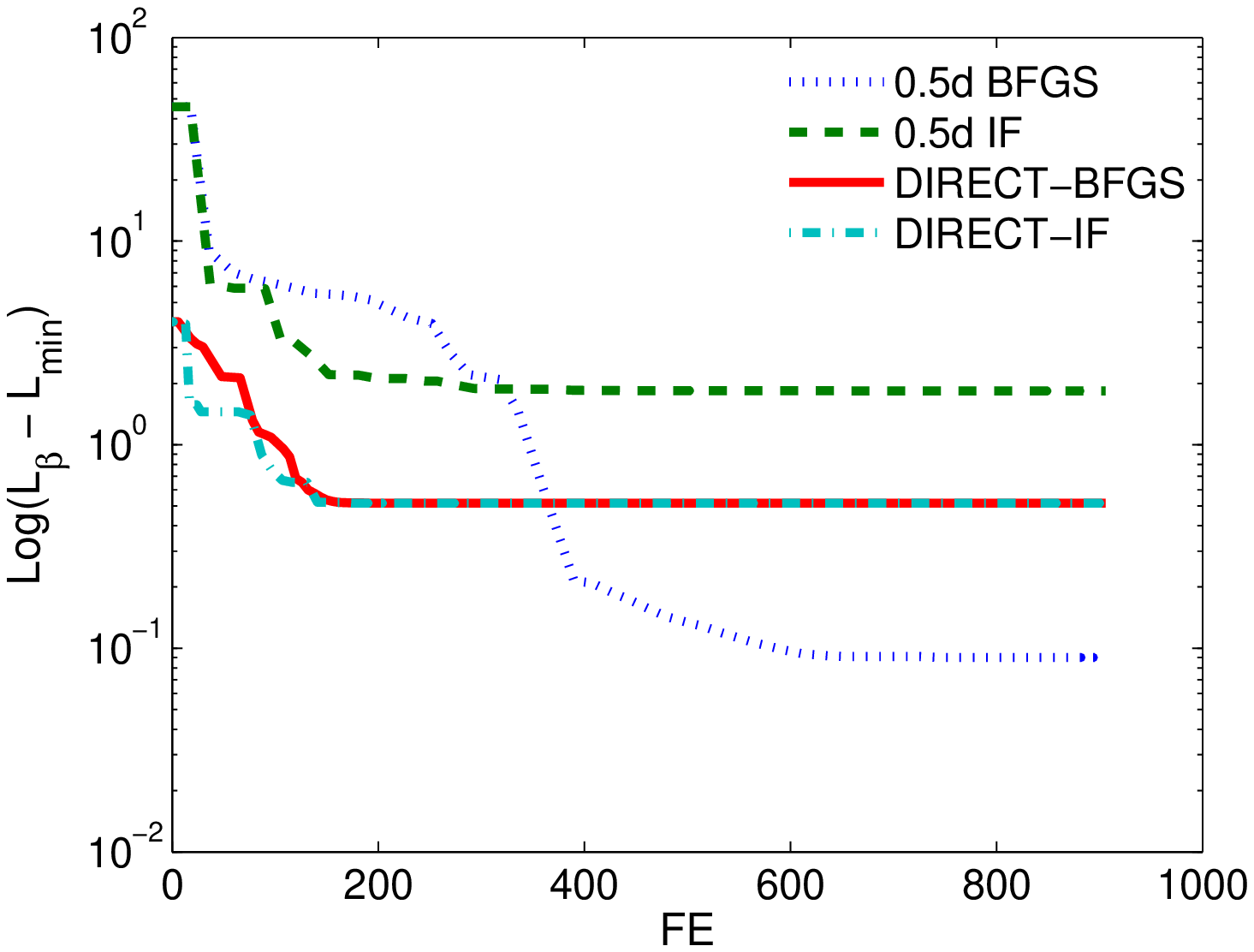}}
\subfigure[Convergence plot.]{\includegraphics[scale=0.40, trim= 1mm 0mm 5mm 5mm, clip=true]{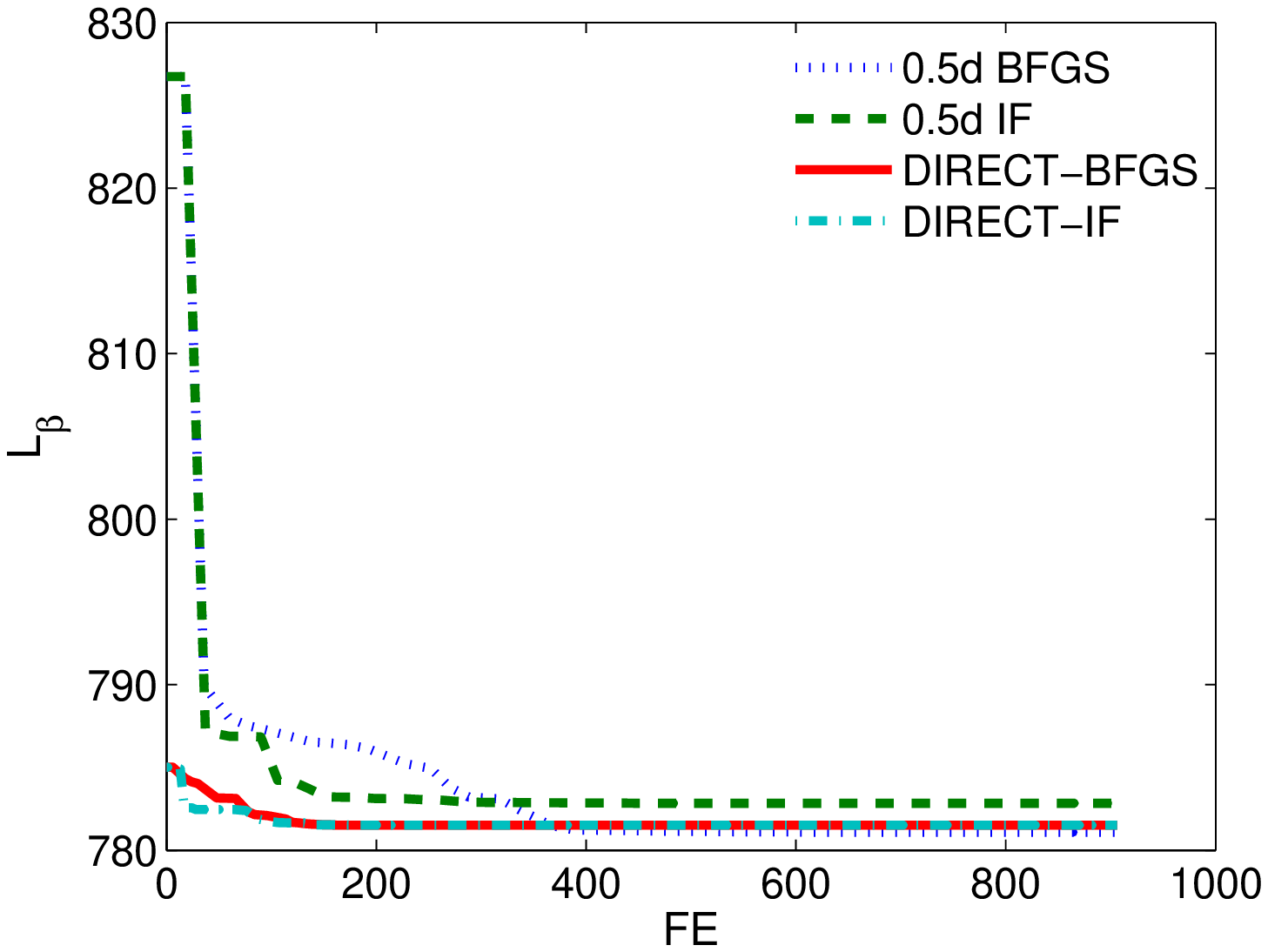}}
\end{center}
\caption[Comparing optimization performance of multi-start and DIRECT methods.]{Semi-log convergence plot and convergence plot comparing the $\mathcal{L}_\beta$ optimization performance of the DIRECT-based techniques and the multi-start clustering techniques.}
\label{schw}

\end{figure}

We ran an additional experiment to determine whether providing additional FEs to the DIRECT-based methods would result in these methods converging to the optimal $\mathcal{L}_\beta$ value found by $\lceil 0.5d \rceil$ BFGS. Figure \ref{schw} compares the optimization performance of the DIRECT-based and $\lceil0.5d \rceil$ multi-start clustering techniques for a single GP model realization of the 5-D Schwefel function. For this particular case, the local search techniques used after DIRECT have been allotted $900$ FEs, which is the number of FEs that were required for $\lceil0.5d \rceil$ BFGS to converge to the optimal $\mathcal{L}_\beta$. From the logarithmic plot (Figure~\ref{schw}(a)), we can see that there is no improvement in the solution found by the DIRECT-BFGS and DIRECT-IF methods after roughly 200 FEs, indicating that allotment of additional FEs provides no benefit to these approaches. We note again, however, that when plotted on regular axes, as in Figure~\ref{schw}(b), the discrepancy in the $\mathcal{L}_\beta$ values returned by the various methods is small.

Figures \ref{performance} and \ref{schw}, show that the DIRECT algorithm is able to determine a more optimal starting position, $\beta^{(0)}$, for initialization of BFGS or IF. This enables us to implement a single run of BFGS or IF, with only a small loss of optimization accuracy. These results were observed for all test functions and support a fundamental conclusion; if the user has significant computational resources at their disposal and if a highly accurate optimal $\mathcal{L}_\beta$ is desired, then the multi-start $\lceil0.5d \rceil$ BFGS technique is preferred. If computational resources are limited, however, then one can use DIRECT-BFGS to obtain comparable GP model fit for a fraction of the computational cost. From these results, we are able to establish an overall performance ranking of each optimization technique, shown in Table \ref{rank}.

\begin{table}[bth]
 \centering
 {
\begin{tabular}{ccc}
 \toprule
\textbf{Rank} & \textbf{Algorithm} & \textbf{\# of starts }\\
\midrule
1 & DIRECT-BFGS & $1$\\
2 & DIRECT-IF & $1$\\
3 & BFGS & $\lceil 0.5d \rceil$\\
4 & IF & $\lceil 0.5d \rceil$\\
5& BFGS & $2d+1$\\
6 & IF & $2d+1$\\
7 &IF-2& $\lceil 0.5d \rceil$\\
\bottomrule
\end{tabular}
}
\caption[Ranks of the performance of each optimization technique.]{Ranks of the optimization techniques based on their overall average performance for the seven test functions.}
\label{rank}

\end{table}

%% file: example.tex
\section{Oil Reservoir Simulator Example}
Determining optimal drilling locations for production and injection wells in an oil reservoir is a problem of considerable industrial interest (see for instance \citet{YDA03, BKWSS06, OD10}). The variables in this problem correspond to positional parameters for each well; in this example, we will consider only vertical wells, each of which can be parameterized by its $(x_1, x_2)$ co-ordinates, representing a grid location in the discrete reservoir model. The well locations serve as input to a computationally expensive complex reservoir simulator -- in our case, the \proglang{Matlab} Reservoir Simulator (MRST)~\citep{LKLNN11,MRST}. The simulator output, along with various economic parameters, are then usually combined to provide the net present value (NPV) of the produced oil. The goal is to determine the configuration of wells that yields the best NPV. 

We consider two problems using a simple 2-D reservoir model based on a 60$\times$50 grid. For the first placement problem, we assume that two injection wells ($\times$) and one production well ($\circ$) have already been drilled at the positions shown in Figure~\ref{reservoir}(a), and the goal is to find the optimal location for the second production well. The NPV surface corresponding to this problem is shown in Figure~\ref{reservoir}(a). One could use an expected improvement based sequential design scheme \citep{Jon} for finding this optimal location; the key component in such a sequential optimization is to efficiently emulate (i.e., fit a GP model to) the simulator response after every iteration of this sequential procedure. In this paper, we focus on this first step of fitting a GP model-based surrogate to the simulator output. For the second problem, we allow the positions of all four wells to be chosen freely, meaning that the NPV now depends on 8 variables. Both of these problems are variants of one considered in~\citet{HHJ13}.



\subsection{2-D Reservoir Simulator }

Table~\ref{table} compares the performance of three methods for fitting the GP model to the 2-D reservoir simulator: (i) $\lceil 0.5d \rceil$ BFGS, (ii) $2d+1$ multi-start BFGS, and (iii) DIRECT-BFGS. The number of training design points used, $n$, ranges from $20$ to $100$. The performance of each optimization technique is averaged over $25$ simulations. For every value of $n$, the $2d+1$ BFGS technique returns the best $\mathcal{L}_\beta$ value. The relative percent difference of the $\mathcal{L}_\beta$ value returned by the DIRECT-BFGS technique, however, is always less than $0.1\%$. Moreover, in all cases, DIRECT-BFGS returns the smallest RMSPE value and requires anywhere from $37\%$ to $45\%$ fewer FEs than $2d+1$ BFGS, and is therefore the preferred technique.

\begin{table}[h!]\centering
{\footnotesize
\begin{tabular}{lccrccr}
\toprule
    \textbf{Algorithm} & \multicolumn{3}{c}{\textbf{n = 20}} & \multicolumn{3}{c}{\textbf{n = 40}}\\

    &\%$\Delta \mathcal{L}_\beta$& \%$\Delta$RMSPE & FE
     & \%$\Delta \mathcal{L}_\beta$& \%$\Delta$RMSPE & FE  \\
     \cmidrule(r){2-4} \cmidrule{5-7}

$\lceil 0.5d \rceil$  BFGS  &0.110& $2.461 $& \underline{435}&2.153& $41.602$ & 452\\
$2d+1$  BFGS   &$-$& $2.461$ & 691&$-$& $3.359$& 721 \\
DIRECT-BFGS& $0.029$& $-$& \underline{435}  &0.004& $-$ & \underline{440}\\

\midrule
     & \multicolumn{3}{c}{\textbf{n =  80}} & \multicolumn{3}{c}{\textbf{n = 100}}\\

    & \%$\Delta \mathcal{L}_\beta$&  \%$\Delta$RMSPE & FE
     & \%$\Delta \mathcal{L}_\beta$&  \%$\Delta$RMSPE & FE  \\
    \cmidrule(r){2-4} \cmidrule{5-7}
$\lceil 0.5d \rceil$ BFGS   &2.757& $47.734$ & 461 &2.648& $22.459$  & 465\\
$2d+1$ BFGS  &$-$& $- $ & 756&$-$& $3.476$ & 822\\
DIRECT-BFGS&$-$& $- $ & \underline{443}&0.072& $-$ & \underline{459}\\
    \bottomrule
\end{tabular}
}
\caption{Performance comparison of different $\mathcal{L}_\beta$ optimization methods for the 2-D reservoir simulator. Dashed values in the \%$\Delta \mathcal{L}_\beta$ and  \%$\Delta$RMSPE columns indicate that the best overall value was found by this algorithm. Underlined values in the FE column indicate the smallest number of FEs required by any algorithm.}
\label{table}
\end{table}

\begin{figure}[h!]\centering
\subfigure[$n=20$]{\includegraphics[height=45mm,trim= 1mm 0mm 5mm 5mm, clip=true ]{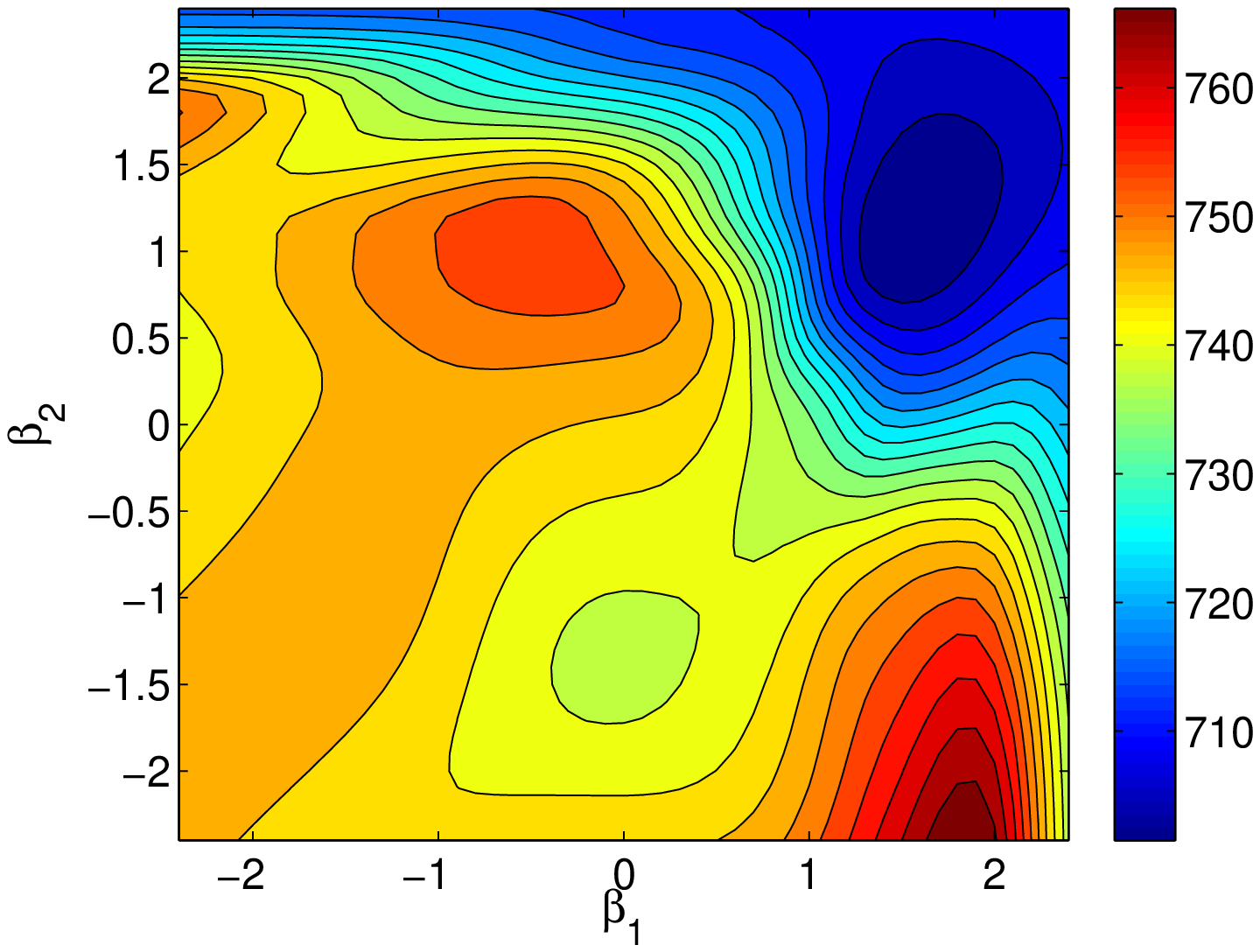}}
\subfigure[$n=80$]{\includegraphics[height=45mm,trim= 1mm 0mm 5mm 5mm, clip=true]{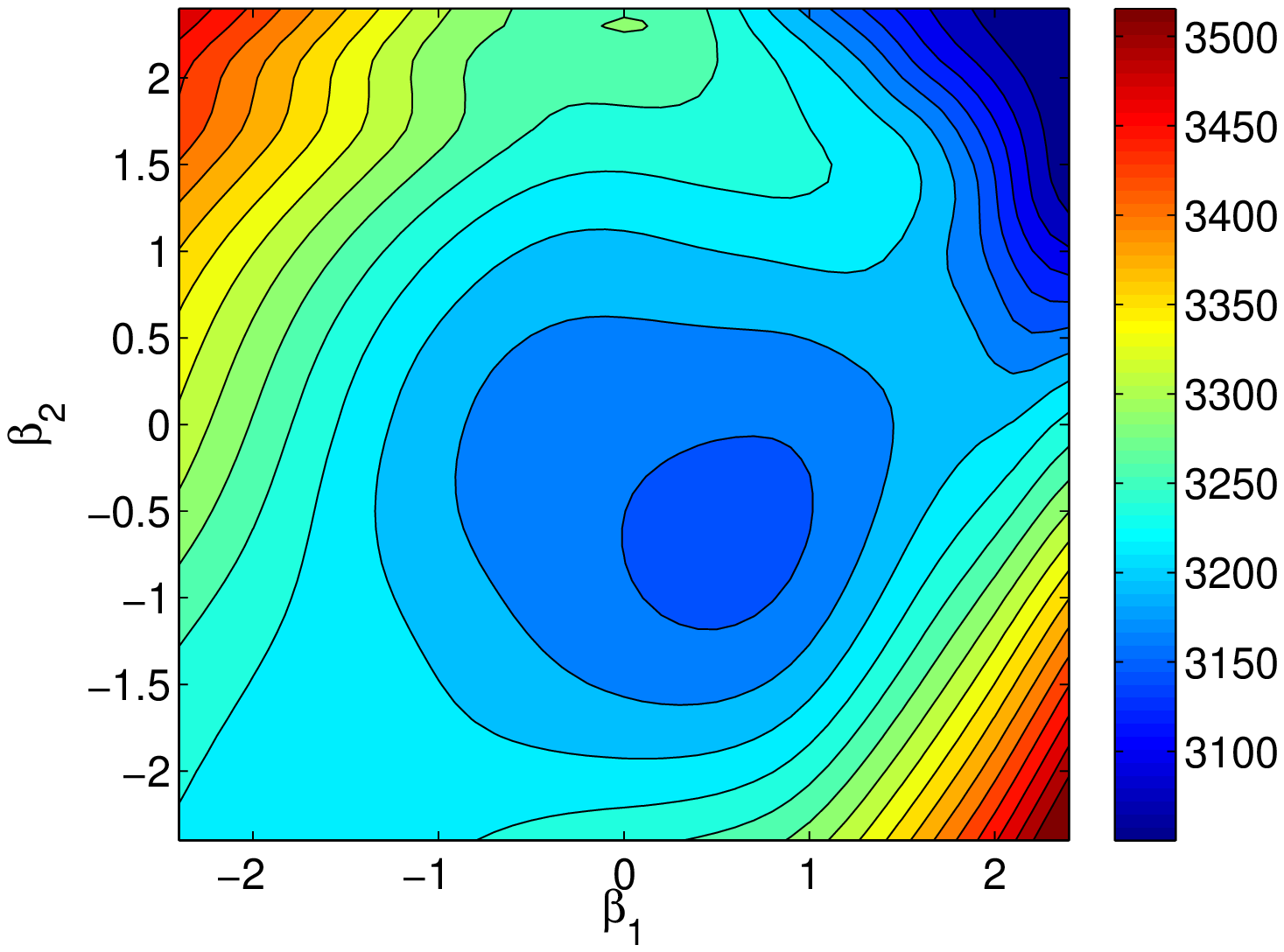}}
\caption{Comparing $\mathcal{L}_\beta$ surfaces of the 2-D reservoir simulator for varying design points.} \label{like_surf}
\end{figure}

A close look on one realization (see Figure \ref{like_surf}) reveals that the likelihood surface changes significanly as the number of design points increases. In particular, when $n=80$, a local optimum with a large basin of attraction is created near the center of the $S^\beta$ domain. As a result, the $\lceil 0.5d \rceil$ multi-start BFGS technique, which is a single-start technique when $d=2$, converges to this sub-optimal point when initialized in that region. DIRECT, on the other hand, is able to determine an starting point that is significantly closer to the global optimum (located in the top right corner) and thus avoids converging to the sub-optimal point. This explains why the $\mathcal{L}_\beta$ and corresponding RMPSE values determined by $\lceil 0.5d\rceil$ BFGS are significantly higher than those values determined by $2d+1$ BFGS and DIRECT-BFGS, when $n \geq 40$.


\begin{figure}[h!]\centering
\subfigure[True simulator output.] {\includegraphics[height=45mm,trim= 1mm 0mm 5mm 5mm, clip=true]{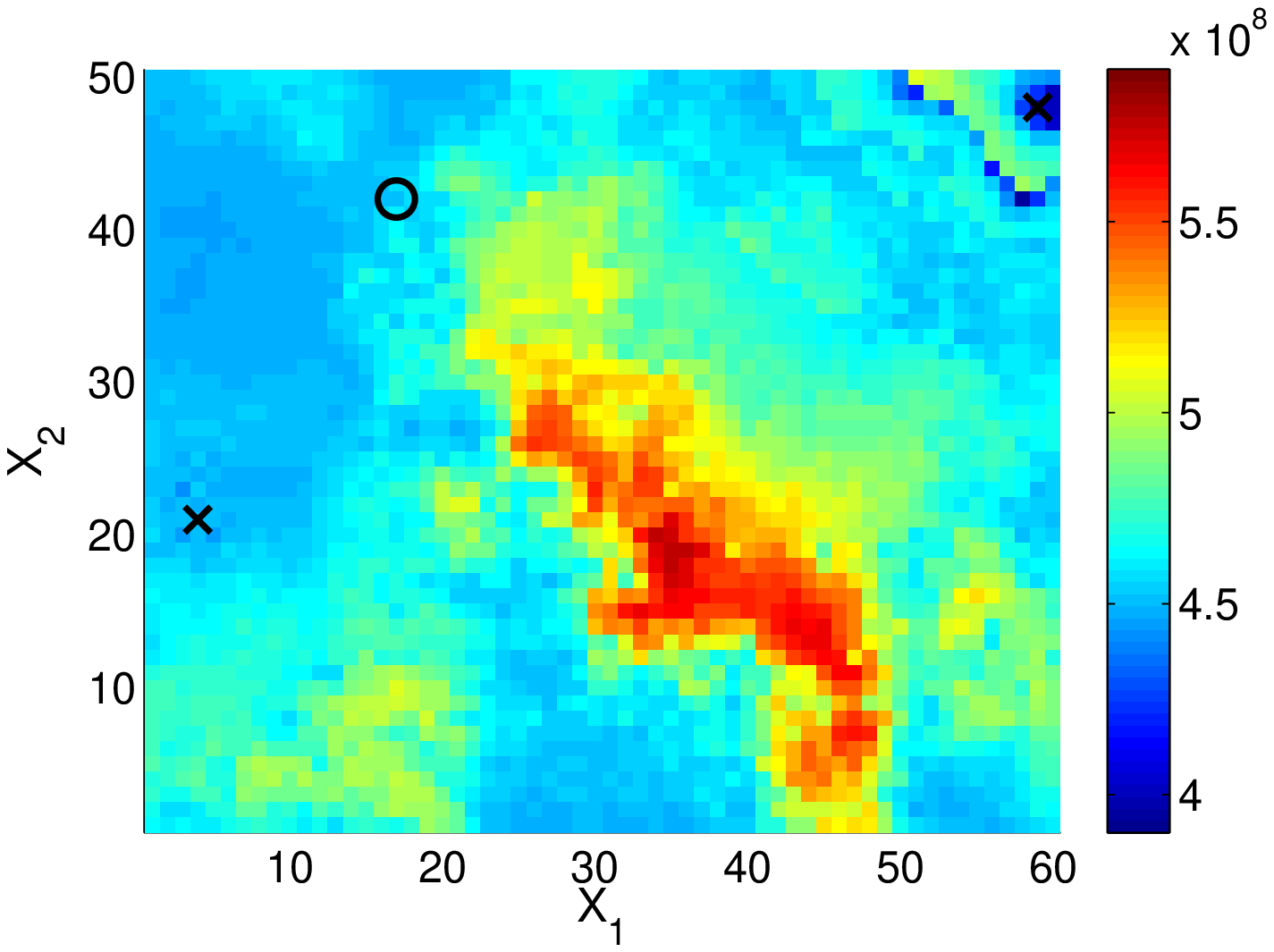}}
\subfigure[$\lceil 0.5d \rceil$ BFGS.] {\includegraphics[height=45mm,trim= 1mm 0mm 5mm 5mm, clip=true]{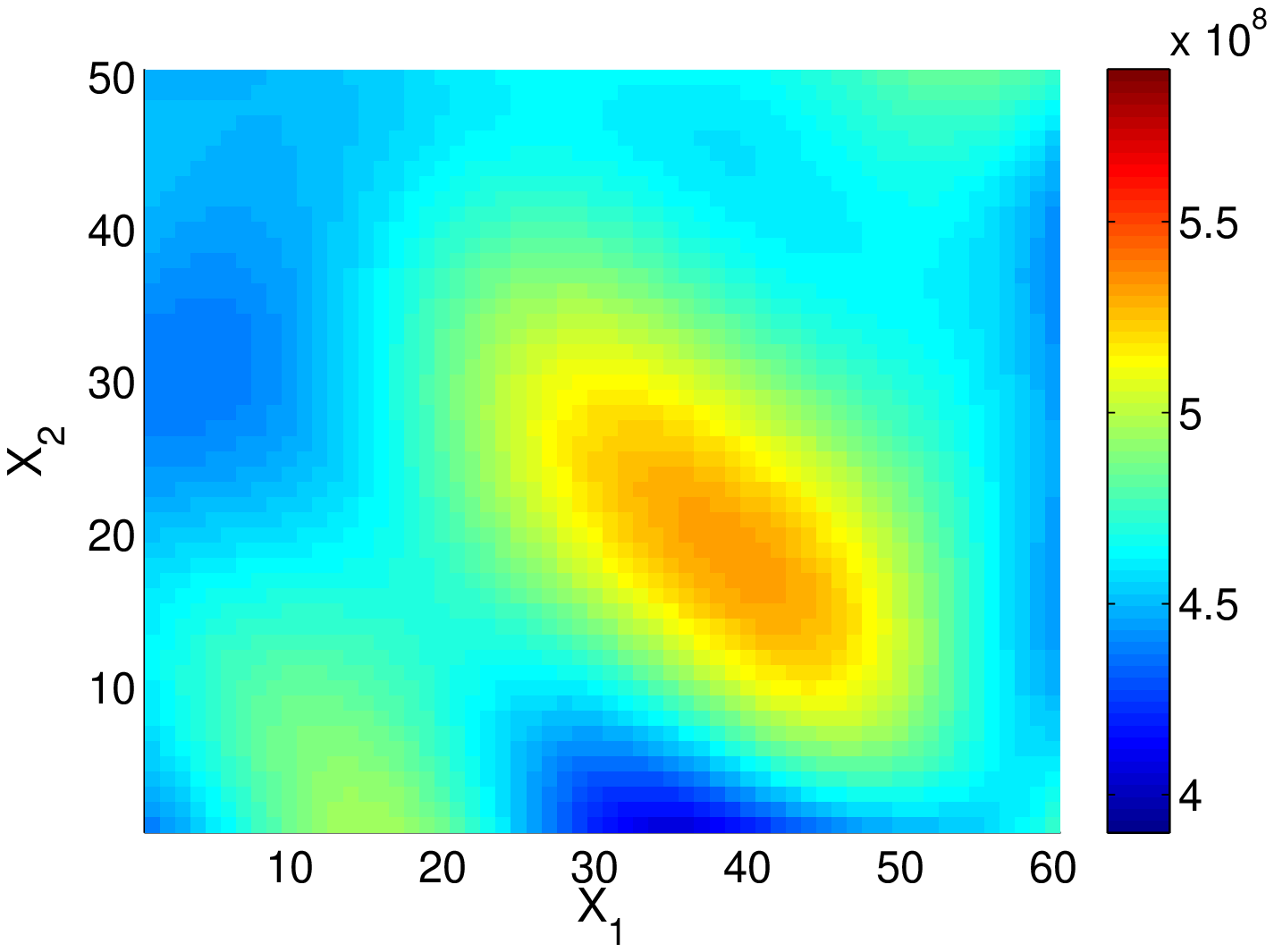}}
\subfigure[$2d+1$ BFGS.] {\includegraphics[height=45mm,trim= 1mm 0mm 5mm 5mm, clip=true]{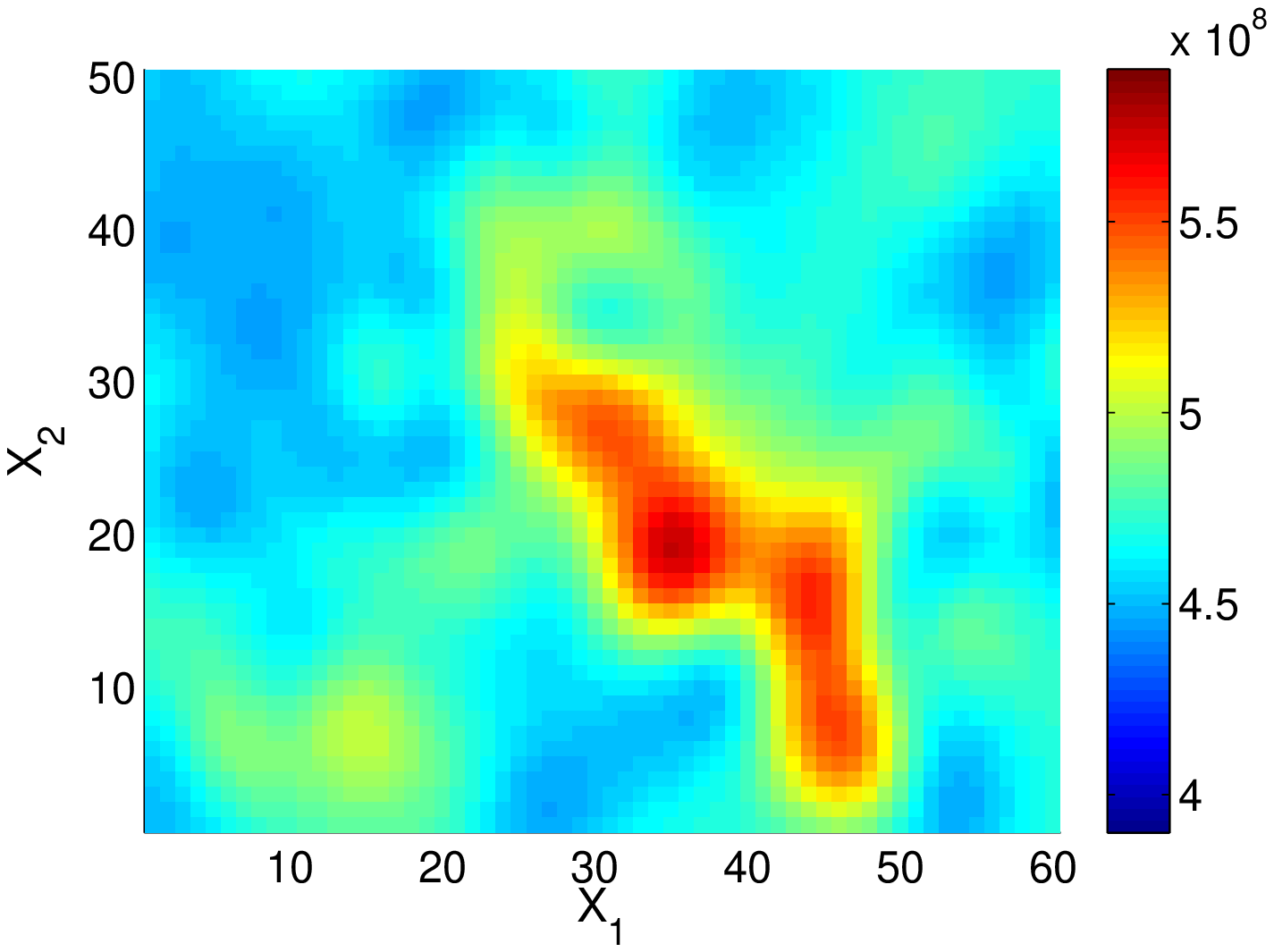}}
\subfigure[DIRECT-BFGS.] {\includegraphics[height=45mm,trim= 1mm 0mm 5mm 5mm, clip=true]{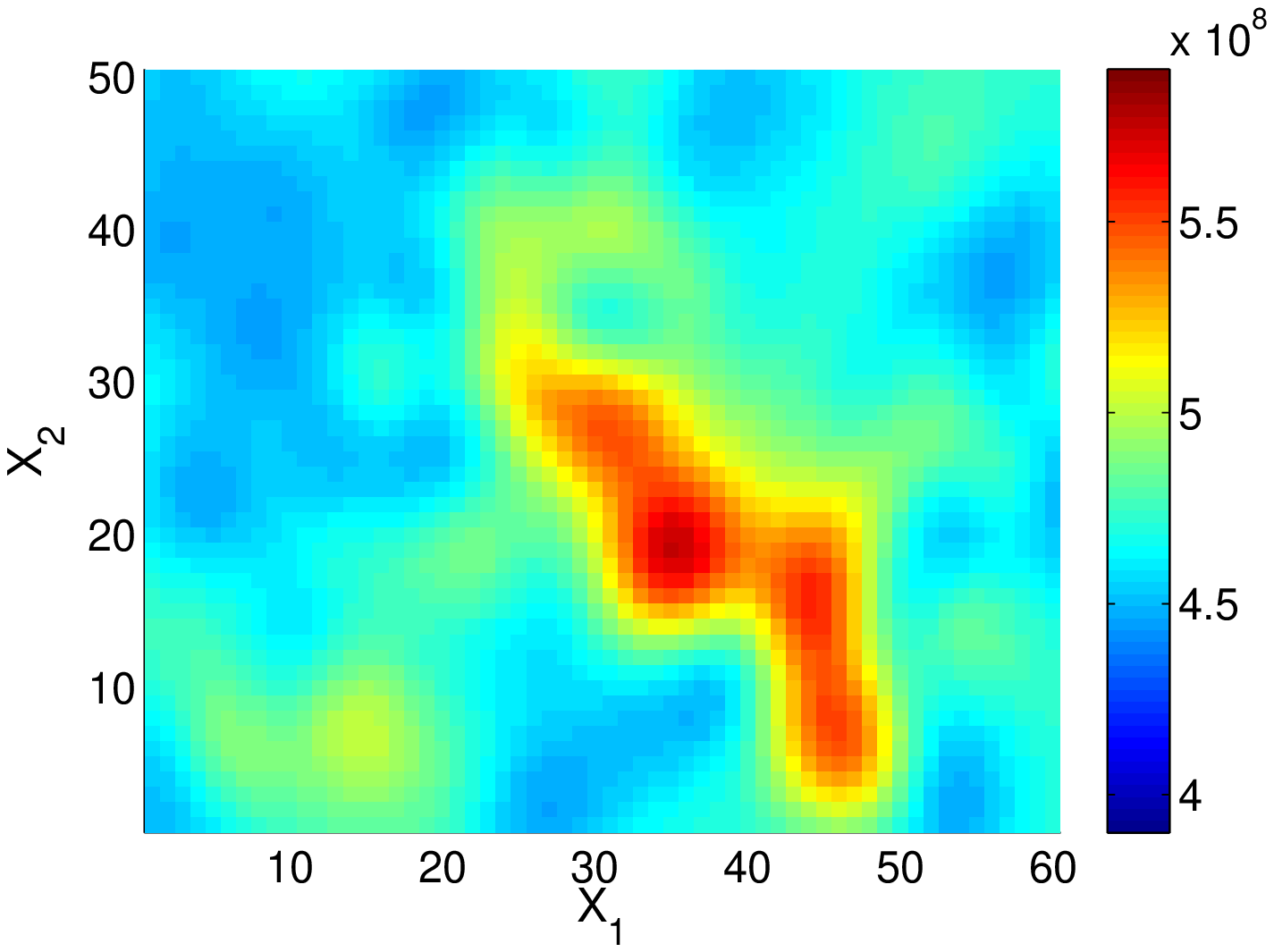}}
\caption{True 2-D reservoir simulator output and GP fit surface obtained through $\mathcal{L}_\beta$ optimization using the $\lceil 0.5d \rceil$ BFGS, $2d+1$ BFGS and DIRECT-BFGS technique, with $n=100$. The locations of the two existing injection ($\times$) wells and single production ($\circ$) well are shown in plot (a).} \label{reservoir}
\end{figure}

Figure \ref{reservoir} shows the true simulator output and an example of the GP models that were found using each of the three optimization techniques, with $n=100$ design points. The $2d+1$ BFGS and DIRECT-BFGS approaches provide near identical GP model predictions, with DIRECT-BFGS  requiring a fraction of the computational cost.  As mentioned, the $\lceil 0.5d \rceil$ BFGS technique converges to a sub-optimal $\beta$-parameter, and as a result the GP model quality is significantly worse. In general we observe that for small dimensional simulators, DIRECT-BFGS outperforms the $\lceil 0.5d \rceil$ multi-start BFGS technique in terms of both optimization accuracy and efficiency (Tables \ref{perform} and \ref{table}, Figure \ref{reservoir}). The $2d+1$ multi-start BFGS method often provides a more accurate GP model fit for any number of design points, but requires up to $80\%$ more FEs than both DIRECT-BFGS and $\lceil 0.5d \rceil$ BFGS for a 2-D function. This example shows that DIRECT-BFGS provides an efficient alternative to the $2d+1$ multi-start BFGS approach, without sacrificing model accuracy.

\subsection{ 8-D Reservoir Simulator}
In our second example, we fit a GP model to the oil reservoir simulator with $8$ variables (the positions of all four wells), again using $\lceil 0.5d \rceil$ BFGS, $2d+1$ BFGS and DIRECT-BFGS for $\mathcal{L}_\beta$ optimization. The GP model is initialized using both $10d$ and $20d$ design points, and predicts for $100d$ points of unknown function value. The performance of each of the three optimization techniques is averaged over $25$ simulations.

Table \ref{8dr} compares the $\mathcal{L}_\beta$ optimization and resulting GP model fitting performance for each of the three techniques. Again, we observe that the $2d+1$ BFGS technique converges to the best $\mathcal{L}_\beta$ value on average. Nonetheless, this method requires roughly $12000$ FEs, which is more than $3$ times the number of FEs required by $\lceil 0.5d \rceil$ BFGS and almost $6$ times the number of FEs required by DIRECT-BFGS. Moreover, DIRECT-BFGS, on average, provides the highest quality GP model, as measured by the RMSPE value, despite converging to a slightly sub-optimal $\mathcal{L}_\beta$ value. The results obtained in fitting the GP model to a true reservoir simulator provide evidence that, in practice, one can employ the single start DIRECT-BFGS optimization technique to greatly increase the efficiency of the GP model fitting procedure, without compromising the quality of the model.

\begin{singlespace}
\begin{table}[h!]
 \centering
 {\footnotesize
 \begin{tabular}{lccrccr}
\toprule
    \textbf{Algorithm} & \multicolumn{3}{c}{\textbf{n = 80}} & \multicolumn{3}{c}{\textbf{n = 160}}\\

    &\%$\Delta \mathcal{L}_\beta$& \%$\Delta$RMSPE & FE
     & \%$\Delta \mathcal{L}_\beta$& \%$\Delta$RMSPE & FE  \\
     \cmidrule(r){2-4} \cmidrule{5-7}

$\lceil 0.5d \rceil$ BFGS   &0.013& $0.711 $& 3706&0.007 & $1.139$ & 3567\\
$2d+1$ BFGS   &$-$& $2.370$ & 11896&$-$& $1.635$& 11893 \\
DIRECT-BFGS& $0.032$& $-$& \underline{2070}  &0.026& $-$ & \underline{2005}\\
\bottomrule
\end{tabular}
}
\caption[Optimization performance for the Oil reservoir simulator.]{Performance comparison of different $\mathcal{L}_\beta$ optimization methods for the 8-D reservoir simulator. Dashed values in the \%$\Delta \mathcal{L}_\beta$ and  \%$\Delta$RMSPE columns indicate that the best overall value was found by this algorithm. Underlined values in the FE column indicate the smallest number of FEs required by any algorithm.} \label{8dr}
\end{table}
\end{singlespace}

%% file: conclusion.tex
\section{Conclusion}
In this paper we have investigated several techniques for efficient optimization of the deviance function, $\mathcal{L}_\beta$, in GP modelling. These techniques provide the foundation for an improved Matlab package, \pkg{GPMfit}. The results obtained from simulated examples and real applications show that using $2d+1$ multi-starts of BFGS is computationally expensive, and that we are able to significantly improve the optimization efficiency by reducing the number of multi-starts to $\lceil 0.5d \rceil$, while maintaining the quality of the GP model in most cases.

Implicit Filtering proves to be slightly less accurate and efficient than BFGS, and therefore is included in the \pkg{GPMfit} solely as a secondary option. The modified multi-start technique IF-2 is generally unreliable; specifically, the slight reduction in computational cost that is gained does not outweigh the reduced accuracy, particularly when more efficient and robust optimization techniques exist.

Replacing the $\lceil 0.5d \rceil$ multi-start technique with the DIRECT optimization algorithm enables us to further reduce the number of starts of BFGS or IF from $\lceil 0.5d \rceil$ to $1$.  After an initial $\beta_0$ value has been determined, the DIRECT-based hybrid techniques require approximately $\frac{1}{\lceil 0.5d \rceil}$ as many function evaluations as the multi-start techniques and provide an almost equally accurate model fit. As a result, the DIRECT-BFGS hybrid technique is the default $\mathcal{L}_\beta$ optimization algorithm used in the \pkg{GPMfit} package. The slightly less efficient DIRECT-IF hybrid technique is also included in the \pkg{GPMfit} package as an additional optimization option, along with the more computationally expensive $2d+1$ and $\lceil 0.5d \rceil$ multi-start BFGS and IF techniques.

\textbf{Future Work:} Training a GP model by expected improvement (EI) is a popular technique for optimization of computationally expensive simulators, in which repeatedly evaluating the simulator by use of traditional optimization routines is inefficient \citep{Jon}. We are currently exploring the use of GP models trained by EI sequential design as an efficient alternative to traditional optimization routines for global optimization of computationally expensive simulators.

\section*{Supplementary Material}
The open source \proglang{Matlab} package \pkg{GPMfit} is available for download on \texttt{SourceForge.net}. See \code{Readme.txt} for detailed instruction. The main functions are \code{model\_fit.m} and \code{predictor\_iterative.m}.

\section*{Acknowledgments}
Haynes, Ranjan and Butler would like to acknowledge the support of the Natural Sciences and Engineering Research Council
of Canada's Discovery Grant and USRA programs.  Humphries was supported by the Research Development Corporation of
Newfoundland and the Atlantic Canada Opportunities Agency.